\documentclass[longauth]{aa} 

\usepackage{graphicx}
\usepackage{txfonts}
\usepackage{subfig}
\usepackage{hyperref}
\hypersetup{hidelinks}
\usepackage{xcolor}

\usepackage{booktabs}
\usepackage{csquotes}
\usepackage{soul}
\defcitealias{Simionescu2017}{S17}
\defcitealias{Kashibadze2020}{K20}
\defcitealias{Cantiello24}{C24}
\defcitealias{Tonnesen10}{T10}

\newcommand{\HI}{\ion{H}{i}}

\newcommand{\Htwo}{\ensuremath{\mathrm{H}_2}}
\begin{document}

   \title{A Virgo Environmental Survey Tracing Ionised Gas Emission (VESTIGE). XXIII. NGC~4569 past orbit reconstruction with \textsc{ORFEO}}

   \titlerunning{VESTIGE  XXIII: NGC 4569 past orbit reconstruction}
   \authorrunning{E.Sarpa et al.}

\author{
    E. Sarpa\thanks{\email{elena.sarpa@inaf.it}}\inst{1,2},
    L. Pizzuti\inst{3},
    A. Boselli\inst{4,5},
    M. Fossati\inst{3,6},
    M. Sun\inst{7},
    E. S. Mangola\inst{4},
    B. Epinat\inst{4,8},
    G. Hensler\inst{9},
    G. De Lucia\inst{1,2},
    L. Cortese\inst{10},
    S. Martocchia\inst{4},
    M. Fumagalli\inst{3,1},
    P. C\^ot\'e\inst{11},
    J.-C. Cuillandre\inst{12},
    L. Ferrarese\inst{11},
    S. Gwyn\inst{11},
    M. Boquien\inst{13},
    F. Renaud\inst{14,15},
    M. A. Taylor\inst{16},
    A. Loni\inst{17},
    P. J\'achym\inst{18},
    P. Amram\inst{4}
}

\institute{
    INAF -- Osservatorio Astronomico di Trieste,
    Via G. B. Tiepolo 11, I-34143 Trieste, Italy
    \and
    IFPU -- Institute for Fundamental Physics of the Universe,
    Via Beirut 2, I-34151 Trieste, Italy
    \and
    Dipartimento di Fisica ``Giuseppe Occhialini'',
    Universit\`a degli Studi di Milano-Bicocca,
    Piazza della Scienza 3, I-20126 Milano, Italy
    \and
    Aix Marseille Universit\'e, CNRS, CNES,
    LAM (Laboratoire d'Astrophysique de Marseille),
    38 rue Fr\'ed\'eric Joliot-Curie, 13388 Marseille, France
    \and
    INAF -- Osservatorio Astronomico di Cagliari,
    Via della Scienza 5, I-09047 Selargius (CA), Italy
    \and
    INAF -- Osservatorio Astronomico di Brera,
    Via Brera 28, I-20121 Milano, Italy
    \and
    Department of Physics and Astronomy,
    University of Alabama in Huntsville,
    301 Sparkman Drive NW, Huntsville, AL 35899, USA
    \and
    Canada-France-Hawaii Telescope,
    65-1238 Mamalahoa Highway,
    Kamuela, HI 96743, USA
    \and
    Department of Astrophysics, University of Vienna,
    T\"urkenschanzstrasse 17, 1180 Vienna, Austria
    \and
    International Centre for Radio Astronomy Research (ICRAR),
    The University of Western Australia,
    35 Stirling Highway, Crawley, WA 6009, Australia
    \and
    National Research Council of Canada,
    Herzberg Astronomy and Astrophysics Research Centre,
    5071 West Saanich Road, Victoria, BC V9E 2E7, Canada
    \and
    Universit\'e Paris-Saclay, Universit\'e Paris Cit\'e,
    CEA, CNRS, AIM,
    91191 Gif-sur-Yvette, France
    \and
    Universit\'e C\^ote d'Azur,
    Observatoire de la C\^ote d'Azur,
    CNRS, Laboratoire Lagrange,
    Boulevard de l'Observatoire,
    BP 4229, F-06304 Nice Cedex 4, France
    \and
    Observatoire Astronomique de Strasbourg,
    Universit\'e de Strasbourg, CNRS UMR 7550,
    11 rue de l'Universit\'e, F-67000 Strasbourg, France
    \and
    University of Strasbourg Institute for Advanced Study,
    5 all\'ee du G\'en\'eral Rouvillois,
    F-67083 Strasbourg, France
    \and
    Department of Physics and Astronomy,
    University of Calgary,
    2500 University Drive NW,
    Calgary, AB T2N 1N4, Canada
    \and
    INAF -- Osservatorio Astronomico di Capodimonte,
    Salita Moiariello 16, I-80131 Napoli, Italy
    \and
    Astronomical Institute, Czech Academy of Sciences,
    Bo\v{c}n\'i II 1401, 14100 Prague, Czechia
}

\date{Received XXXX; accepted XXX}

  \abstract{

Extended ionised gas tails can preserve a dynamical record of a galaxy's recent interaction with the intracluster medium (ICM). However, recovering that history from the observed tail requires solving an inverse dynamical problem tied to the physical conditions for gas removal. We present the first application of \textsc{ORFEO} (Orbits Reconstruction For Emitting Objects) to NGC~4569, using the sky positions and line-of-sight (LOS) velocities of the galaxy and 94 H\(\alpha\)-emitting regions to reconstruct its recent orbit and tail chronology. The model combines these data with gas surface-density constraints, a spherical Virgo Cluster~A gravitational potential derived with \textsc{MG-MAMPOSSt}, and an X-ray-based ICM density profile. For an assumed LOS distance, it infers the galaxy orbit and corresponding ram-pressure history. At the fiducial distance \(d_0=16.5\,\mathrm{Mpc}\), \textsc{ORFEO} recovers a coherent phase-space sequence with a dynamical tail age \(T_\mathrm{tail}=250^{+37}_{-19}\,\mathrm{Myr}\). The ram-pressure maximum occurred at \(t_{P_\mathrm{ram},\mathrm{peak}}=-163^{+11}_{-15}\,\mathrm{Myr}\), close to the nearest approach to the Virgo Cluster~A centre at \(305^{+19}_{-25}\,\mathrm{kpc}\), placing NGC~4569 after peak ram pressure. Scanning the LOS distance does not select a unique solution but separates comparably fitting reconstructions into a low-distance post-peak branch and a high-distance pre-peak branch. Comparison with an independent stripping threshold at the observed disc truncation radius favours the low-distance solutions and supports a post-stripping interpretation. On this branch, \textsc{ORFEO} also recovers the mostly west-to-east projected motion inferred from the H\(\alpha\) morphology and is consistent with previous constraints on the interaction chronology. Requiring only \(\sim10\)--\(11\,\mathrm{min}\) of wall-clock time for NGC~4569, \textsc{ORFEO} provides a computationally inexpensive inverse framework for deriving orbital, tail, and ram-pressure histories that can guide tailored simulations and be extended to larger samples of stripped galaxies.

}


\maketitle
%
\section{Introduction}

Galaxies in dense environments differ systematically from their field counterparts. Dense environments contain a higher fraction of early-type galaxies 
\citep{Dressler80}, while cluster spirals have lower atomic gas content \citep{Haynes84,Cayatte90,Solanes01,Gavazzi05}; H\,{\sc i}-deficient systems can also show reduced molecular-gas reservoirs \citep{Fumagalli09,Boselli14b} and star-formation activity \citep{Kennicutt83,Gavazzi98,Lewis2002,Gomez03,Boselli14c}. These trends demonstrate the important role of environment in regulating galaxy evolution \citep[see e.g.][]{Peng10}, but the present-day state of a galaxy does not by itself identify the physical process or interaction history that produced it.

Several mechanisms can drive this transformation \citep[see reviews by][]{BG06,BG14,Cortese21}, and their different physical action gives rise to different characteristic timescales. Gravitational interactions with nearby galaxies and with the global cluster potential can perturb both the stellar and gaseous components \citep{Merritt83,Byrd90,Moore98}; repeated high-speed encounters in particular can produce cumulative transformations over several Gyr \citep{Moore98,BG06}. Starvation instead suppresses the replenishment of the cold interstellar medium \citep{Larson80}, so star formation declines gradually as the existing gas reservoir is consumed over several Gyr \citep{Boselli06,Boselli14c}. Ram-pressure stripping (RPS), by contrast, acts directly on the interstellar medium when the pressure exerted by the intracluster medium (ICM) overcomes the gravitational restoring force of the galaxy \citep{Gunn72,Boselli22}, and can remove gas on sub-Gyr timescales, producing truncated star-forming discs and extended gaseous tails \citep{Koopmann04,Boselli06,Crowl08,Cortese12,Boselli14c,Boselli16b,Fossati18,Boselli22,Boselli23a}.\newline
\indent The timing and effectiveness of environmental processing also depend on the accretion history and orbit of a galaxy. Environmental effects can begin before cluster infall, through pre-processing in groups \citep{McGee09,DeLucia12,Fossati19} or suppression of gas replenishment in the cluster outskirts \citep{Bahe13,Ayromlou21}, and continue after cluster entry through interactions with other galaxies and with the global cluster potential. RPS is particularly tied to the galaxy trajectory: it becomes more effective where the galaxy encounters denser ICM and high relative velocities, and therefore changes in strength and timing as these conditions vary along the orbit \citep{Gunn72,Boselli22}.

The present-day properties of a galaxy therefore do not uniquely reveal the interaction history that produced them \citep[see e.g.][]{Cortese21}. Distinguishing between these environmental pathways requires information that cannot in general be recovered uniquely from the present-day galaxy alone. The central questions are therefore when, and over what timescale, the material now observed in the tail was removed, where the galaxy was along its orbit, and what environmental conditions it encountered.

Extended gaseous tails offer a direct opportunity to address this temporal and orbital problem. Because stripped material can remain observable outside the galaxy disc, its present-day spatial and kinematic structure can retain information on its evolution after removal. In favourable systems, resolved positions and velocities along a tail may therefore preserve a dynamical record of different stages of the recent interaction \citep{Fumagalli14,Luo23,Sun2026}. The Virgo Environmental Survey Tracing Ionised Gas Emission (VESTIGE), a deep narrow-band H\(\alpha\) imaging survey of the Virgo cluster carried out at the Canada--France--Hawaii Telescope (CFHT; \citealt{Boselli18a}), has revealed several galaxies with extended ionised-gas structures associated with ongoing or recent stripping.

NGC~4569 provides a particularly well-suited benchmark. Its ionised-gas tail was originally detected in H\(\alpha\) out to \(\simeq80\,\mathrm{kpc}\) \citep{Boselli16a}, while deeper observations now trace low-surface-brightness emission to at least \(\simeq230\,\mathrm{kpc}\) \citep{Sun2026}. New spectroscopy provides LOS velocities for 94 H\(\alpha\)-emitting regions along the galaxy and tail \citep{Sun2026}, revealing a coherent kinematic sequence and turning this extended structure from a predominantly morphological signature into a resolved spatial and kinematic record of the recent interaction. Previous modelling has also yielded different characteristic epochs for the interaction: \citet{Vollmer04} favoured a peak in ram pressure roughly \(300\,\mathrm{Myr}\) ago, whereas \citet{Boselli06} inferred a substantially more recent stripping event, of order \(100\,\mathrm{Myr}\) ago. NGC~4569 therefore provides both the observational leverage needed to attempt a dynamical reconstruction and independent estimates against which its inferred chronology can be assessed.

Previous studies have constrained environmental histories using complementary observables. Spatially resolved stellar populations and star-formation histories provide estimates of quenching timescales \citep{Boselli06,Crowl08}; tail morphology and orientation constrain the projected direction of motion and, with additional assumptions, aspects of the galaxy orbit \citep{Randall08, McPartland2016}; cluster projected phase space provides statistical information on infall, orbital, and stripping stages \citep{Jaffe2018}; resolved gas spectroscopy traces the dynamics of stripped material \citep{Fumagalli14,Luo23,Sun2026}; and the surviving gas distribution constrains the strength and evolutionary stage of stripping \citep{koppen18}. At the level of individual systems, tailored dynamical and hydrodynamical models combine morphology and kinematics to constrain stripping epochs, wind geometry, ram-pressure histories, and orbital properties \citep{Vollmer04,Merluzzi13,Boselli23c,Vollmer24}. These tailored models are predominantly forward: candidate interaction histories are specified or explored, evolved, and their predicted observables compared with the data.

Here we introduce the complementary inverse approach implemented in \textsc{ORFEO} (Orbits Reconstruction For Emitting Objects). Rather than starting from a prescribed interaction history and asking whether it reproduces the observed tail, \textsc{ORFEO} uses the resolved positions and LOS velocities of the individual tail tracers to directly constrain the recent galaxy trajectory and tail chronology\footnote{Throughout this work, the tail chronology denotes the dynamical ordering and lookback-time sequence assigned to the observed tail tracers.}, without prescribing the interaction history that generated the observed structure. To our knowledge, this is the first attempt to reconstruct the recent dynamical history of a cluster galaxy by directly inverting the resolved spatial and kinematic sequence encoded in its ionised-gas tail.

In this first application of \textsc{ORFEO}, we reconstruct the recent trajectory of NGC~4569 and the age sequence of its observed H\(\alpha\) tail over a range of assumed LOS distances. For each reconstruction, the corresponding ram-pressure history is then derived from the inferred trajectory and used to assess whether RPS provides a physically consistent explanation for the observed stripping. As a computationally inexpensive complement to tailored simulations, this inverse approach may ultimately make time-resolved interaction histories accessible for larger samples of stripped galaxies.

The data and \textsc{ORFEO} framework are presented in Sects.~\ref{Sec:Data} and \ref{Sec:Orfeo}, respectively. The reconstruction and its dependence on LOS distance are presented in Sect.~\ref{Sec:results}. The physical assessment of the solutions and comparison with previous models are discussed in Sect.~\ref{sec:discussion}. We summarise our conclusions in Sect.~\ref{sec:conclusions}.


\section{Data}
\label{Sec:Data}

For NGC~4569, we use the complete catalogue of 94 robust H\(\alpha\)-emitting regions from \citet{Sun2026} as direct constraints on the reconstruction. The regions trace the extended ionised-gas tail identified by VESTIGE, a deep narrow-band H\(\alpha\) survey of the Virgo cluster carried out with MegaCam at CFHT \citep{Boselli16a,Boselli18a,Boselli19}. At a resolution of \(\sim3\,\arcsec\), the VESTIGE observations reach a \(1\sigma\) limiting surface brightness of \(\Sigma(\mathrm{H}\alpha)\simeq1.5\times10^{-18}\,\mathrm{erg\,s^{-1}\,cm^{-2}\,arcsec^{-2}}\). They trace low-surface-brightness emission to a projected distance of \(\simeq230\,\mathrm{kpc}\), extending well beyond the \(\sim80\,\mathrm{kpc}\) bright tail identified in the initial observations \citep{Boselli16a,Sun2026}.

\begin{figure}
\centering
\includegraphics[width=1.\columnwidth]{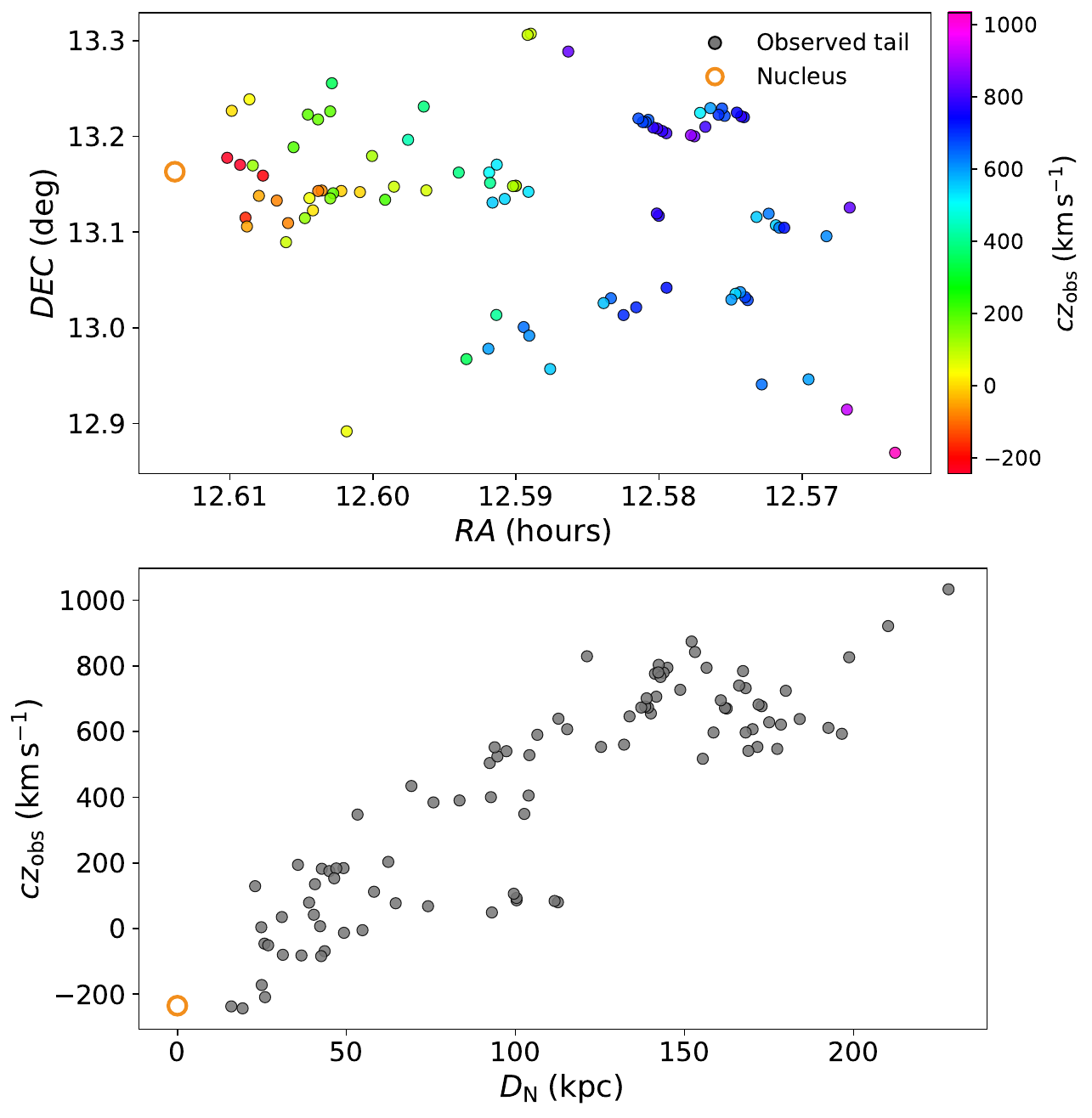}
\caption{Observed position and LOS velocity of NGC~4569 and its ionised-gas tail. \emph{Top panel}: right ascension and declination of NGC~4569 and of the 94 H\(\alpha\)-emitting regions used in the reconstruction. Points are coloured according to the measured LOS velocity.} \emph{Bottom panel}: observed recession velocity, \(cz_\mathrm{obs}\), of the H\(\alpha\)-emitting regions as a function of projected distance from the galaxy nucleus, \(D_\mathrm{N}\), computed in the parallel-plane approximation. The orange circle marks the galaxy nucleus.
\label{fig:NGC4569_data}
\end{figure}
\subsection{NGC~4569}
\label{Sec:data_NGC}
The spectroscopic measurements used in the reconstruction were obtained with the Binospec multi-object spectrograph on the 6.5-m MMT telescope using the 600 lines mm\(^{-1}\) grating. The spectra have a typical wavelength coverage of 4900--7200\,\text{\AA}, a dispersion of \(0.62\,\text{\AA}\,\mathrm{pixel}^{-1}\), and a resolving power \(R\simeq2800\) for \(1\arcsec\) slits. The measured velocities have a median statistical uncertainty of \(5\,\mathrm{km\,s^{-1}}\) and a systematic uncertainty of \(6\)--\(10\,\mathrm{km\,s^{-1}}\).
Each of the 94 regions provides the three observables entering the likelihood: right ascension (RA), declination (DEC), and the observed spectroscopic redshift expressed in velocity units, \(cz_\mathrm{obs}\). The H\(\alpha\)-emitting regions span \(cz_\mathrm{obs}\) from \(-243\) to \(+1033\,\mathrm{km\,s^{-1}}\) and projected distances from approximately \(16\) to \(228\,\mathrm{kpc}\) for the fiducial distance \(d_0=16.5\,\mathrm{Mpc}\). The nucleus of NGC~4569, which anchors the present-day galaxy position and systemic spectroscopic redshift, is located at \((\mathrm{RA}_0,\mathrm{DEC}_0)=(12.61384582\,\mathrm{h},13.1631611^\circ)\) and has \(cz_{\mathrm{obs},0}=-235\,\mathrm{km\,s^{-1}}\). Figure~\ref{fig:NGC4569_data} shows the galaxy nucleus and the observables used in the reconstruction, both in the sky plane and in the \(cz_\mathrm{obs}\)--\(D_\mathrm{N}\) projection, where \(D_\mathrm{N}\) is the projected distance from the galaxy nucleus in the parallel-plane approximation. The association of the H\(\alpha\)-emitting
regions with NGC~4569 is supported by their coherent spatial
distribution extending from the galaxy and by the continuous
trend of recession velocity with projected distance, connecting
the tail kinematics to the galaxy's systemic velocity
\citep{Sun2026}.

In addition to the phase-space observables, the dynamical model
of the stripped gas described in Sect.~\ref{sec:Gaseous_tail}
requires a range of effective gas surface densities,
representing the denser material near the disc and the diffuse
gas at the outer end of the modelled tail. We use an observationally motivated range for the minimum effective surface density and a fixed dense-disc reference value for the maximum. We set the minimum effective surface density of the diffuse outer-tail gas using the neutral-gas upper-limit
scale \(N_\mathrm{H}\simeq(2\text{--}5)\times10^{19}\,\mathrm{cm^{-2}}\)
inferred by \citet{Boselli16a} from the dust non-detection,
assuming dust properties and gas-to-dust ratios typical of
nearby massive galaxies.
This scale is comparable to the VIVA \ion{H}{i} sensitivity
\citep{Chung2009}. For the dense disc gas, we adopt a reference surface density \(\Sigma_\mathrm{disc,ref}\simeq10\,M_\odot\,\mathrm{pc^{-2}}\), representative of the observed total-gas surface density near the present-day disc truncation radius, \(R_\mathrm{strip}\simeq4.8\,\mathrm{kpc}\), in the radial profile of NGC~4569 derived by \citet{Boselli06} from \ion{H}{i} and CO observations \citep{Cayatte94,Helfer03,Kenney88}. The observed profile is smoothed to \(1\,\mathrm{kpc}\) resolution and does not provide a formal pointwise uncertainty at this radius. We convert column densities into gas surface densities using \(\Sigma_\mathrm{gas}=\mu m_p N_\mathrm{H}\), where \(m_p\) is the proton mass and \(\mu=1.3\) accounts for helium. The quoted tail upper-limit scale motivates the adopted range for the model minimum, $\Sigma_{\mathrm{tail,min}} \in [4.3,10.9]\times10^{-5}\, \mathrm{g\,cm^{-2}}$. The upper endpoint is fixed to the dense-disc reference value, $\Sigma_{\mathrm{disc,ref}} = 217.4\times10^{-5}\,
\mathrm{g\,cm^{-2}}$. These values set the effective surface-density interval used to model the ram-pressure deceleration of the stripped gas parcels.

\subsection{Virgo Cluster A}
\label{sec:Virgo}

The reconstruction also requires a description of the Virgo Cluster~A environment. Its mass distribution sets the gravitational potential in which NGC~4569 and the stripped gas evolve, while the ICM density profile enters the dynamical treatment of the stripped material and the reconstruction of the ram-pressure history.
We infer the Virgo Cluster~A mass profile from the positions and redshifts of member galaxies selected from the Virgo catalogue of \citet{Boselli14c,Boselli23a}. The parent catalogue contains 868 cluster members within a projected radius \(R\simeq2\,R_{200}\). Membership is based primarily on spectroscopic redshifts, using in order of priority \ion{H}{i} measurements from the ALFALFA survey \citep{Giovanelli05,Haynes18}, targeted \ion{H}{i} observations compiled in the GoldMine database \citep{Gavazzi03}, and optical spectroscopy, mainly from SDSS and NED. We assign galaxies to Virgo substructures following the classification scheme of \citet{Boselli14c,Boselli23a}, which combines sky position and redshift. In regions where a galaxy satisfies the criteria for more than one substructure, we associate it with the smaller structure. Distances to the individual substructures are adopted from \citet{Boselli23b}. Early- and late-type galaxies are separated using the morphological classifications of the Virgo Cluster Catalogue (VCC; \citealt{Binggeli85}), later updated by \citet{Binggeli93}. 
For the present analysis, we restrict the dynamical sample to galaxies associated with Cluster~A, the M87-centred component that dominates the gravitational potential relevant for the orbit of NGC~4569. We adopt the position of M87, \((\mathrm{RA}_\mathrm{cl},\mathrm{DEC}_\mathrm{cl})=(12.516040\,\mathrm{h},12.270700^\circ)\), as the spatial centre and \(d_\mathrm{cl}=16.5\,\mathrm{Mpc}\) as the reference distance of Cluster~A \citep{Mei2007,Cantiello24}. To define the LOS component of this reference frame, we use the observed recession velocity of M87, \(cz_\mathrm{cl}=1307\,\mathrm{km\,s^{-1}}\). Relative to the Hubble flow, this corresponds to \(v_{\parallel,\mathrm{cl}}=cz_\mathrm{cl}-d_\mathrm{cl}H_0=102.5\,\mathrm{km\,s^{-1}}\), for \(H_0=73\,\mathrm{km\,s^{-1}\,Mpc^{-1}}\). Together, these quantities define the M87-centred reference frame used to compute galaxy peculiar velocities and initialise the ORFEO orbit reconstruction. The resulting Cluster~A sample contains \(285\) galaxies with spectroscopic recession velocities in the range \(-730\leq cz\leq3000\,\mathrm{km\,s^{-1}}\), which reflects the adopted membership classification rather than a symmetric velocity cut around \(cz_\mathrm{cl}\). Of these, \(185\) are classified as early types and \(100\) as late types. The dynamical modelling of this sample and the derivation of the NFW potential used by \textsc{ORFEO} are described in Sect.~\ref{sec:manposst}.

To describe the interaction of the stripped gas with the Virgo ICM and to evaluate the ram pressure along the reconstructed trajectory, we adopt a spherically symmetric analytic approximation to the Virgo gas-density profile based on the X-ray analysis of \citet{Simionescu2017}. The profile is centred on M87 and is written in terms of the electron density as \(n_{\mathrm{e,ICM}}(r)=n_{\mathrm{e},0}\left[1+\left(r/r_\mathrm{c}\right)^2\right]^{-3\beta/2}\), with \(n_{\mathrm{e},0}=3.13\times10^{-2}\,\mathrm{cm^{-3}}\), \(r_\mathrm{c}=13\,\mathrm{kpc}\), and \(\beta=0.47\). This profile is fixed throughout the analysis. When computing ram pressure, we convert the electron density into a gas mass density before evaluating \(P_\mathrm{ram}=\rho_\mathrm{ICM}v^2\).

\section{Orbit reconstruction}
\label{Sec:Orfeo}

The \textsc{ORFEO} algorithm reconstructs the orbit of a cluster galaxy by using the phase-space structure of its stripped gas as a dynamical tracer of its recent motion through the cluster environment. For each trial present-day three-dimensional position and velocity in the cluster rest frame, \textsc{ORFEO} integrates the galaxy trajectory in the host-cluster potential. The tail is then constructed by sampling point-like parcels at different lookback times along this trajectory and evolving them to the present under cluster gravity and ram-pressure drag, defining a model that is projected onto sky position and LOS velocity. For a fixed assumed LOS distance, comparison with the observed H\(\alpha\)-emitting regions constrains the two unknown transverse velocity components of the galaxy.
\subsection{Orbit equations}
\label{sec:orbit_equations}

We model the galaxy as a test particle moving in the spherically symmetric gravitational potential of the host cluster, \(\phi(\mathbf{r})=\phi(r)\). The specific energy and specific angular momentum are conserved,
\begin{eqnarray}
e &=& \frac{E}{m} =
\frac{v(t)^2}{2}+\phi(r(t)),
\label{eq:energy_conservation}
\\
\boldsymbol{\ell} &=& \frac{\mathbf{L}}{m} =
\mathbf{r}(t)\times\mathbf{v}(t),
\label{eq:momentum_conservation}
\end{eqnarray}
where \(\mathbf{r}\) and \(\mathbf{v}\) are the position and velocity in the cluster rest frame, \(r=|\mathbf{r}|\), \(v=|\mathbf{v}|\), and \(m\) is the galaxy mass. Using specific quantities removes the explicit mass dependence from the orbit solution.
For each trial state \((\mathbf{r}_0,\mathbf{v}_0)\) at time \(t_0\), \textsc{ORFEO} computes \(e\) and \(\boldsymbol{\ell}\), rotates the coordinate system so that the \(z\)-axis is aligned with \(\boldsymbol{\ell}\), and solves the motion in the \((x,y)\) orbital plane. In this frame, the position is described by the polar coordinates
\(r(t)=\sqrt{x(t)^2+y(t)^2}\) and
\(\theta(t)=\operatorname{atan2}[y(t),x(t)]\).
Defining the effective radial potential
\(u_\mathrm{eff}(r)=\phi(r)+\frac{\ell^2}{2r^2}\), \(\ell=|\boldsymbol{\ell}|\), energy conservation gives the radial evolution as
\begin{equation}
t=t_0+\frac{1}{\sqrt{2}}
\int_{r_0}^{r(t)}
\frac{\mathrm{d}r'}{\sqrt{e-u_\mathrm{eff}(r')}} .
\label{eq:time_evolution}
\end{equation}
For radial orbits, \(\ell=0\) and \(\theta(t)=\theta_0\). For non-radial orbits, \textsc{ORFEO} determines the radial turning points from \(e-u_\mathrm{eff}(r)=0\), distinguishing between unbound trajectories, \(e\geq0\), and bound trajectories, \(e<0\). At each turning point, the sign of the radial velocity is reversed, while the angular coordinate evolves as
\begin{equation}
\theta(t)=\theta_0+
\int_{t_0}^{t}
\frac{\ell}{r(t')^2}\,\mathrm{d}t' .
\label{eq:ang_evolution_unb}
\end{equation}
The trajectory is then transformed back to the original cluster-centred reference frame. 

To reconstruct the past orbit, \textsc{ORFEO} integrates the time-reversed trajectory forward from the present state, using the reversed peculiar velocity. The resulting solution is then mapped back to physical time and velocity, such that \(t=0\) at the present epoch, \(t<0\) in the past, and \(\mathbf{v}(t)\) has the physical velocity sign.
\begin{table*}
    \centering
    \small
    \setlength{\tabcolsep}{3pt}
    \caption{
  Virgo Cluster~A dynamical-model parameters from the \textsc{MG-MAMPOSSt} analysis for the NFW mass model, generalised Osipkov--Merritt anisotropy profile, and projected Hernquist tracer-density profile. Results are shown for the early-type subsample (ET), adopted as the fiducial \textsc{ORFEO} potential, and for the full galaxy sample (ALL), reported for comparison. Here \(r_\nu\) is the tracer scale radius, while \(\mathcal{A}_0\) and \(\mathcal{A}_\infty\) parametrise the anisotropy profile. The final column reports \(\Delta\mathrm{BIC}\), the difference in Bayesian information criterion used to compare the dynamical models.
    }
    \begin{tabular}{c c c c | c c c c c c c}
    \toprule
    \(M(r)\) model & \(\beta(r)\) model & \(N(r)\) model & sample
    & \(r_{200}\) [Mpc]
    & \(M_{200}\) [\(10^{14}\,M_\odot\)]
    & \(R_\mathrm{s}\) [Mpc]
    & \(r_\nu\) [Mpc]
    & \(\mathcal{A}_\infty\)
    & \(\mathcal{A}_0\)
    & \(\Delta\)BIC \\
    \midrule
    NFW & gOM & pHer & ET
    & \(1.63^{+0.32}_{-0.21}\)
    & \(5.4^{+2.9}_{-2.3}\)
    & \(0.36^{+0.32}_{-0.07}\)
    & \(0.50^{+0.12}_{-0.06}\)
    & \(1.70^{+1.86}_{-1.23}\)
    & \(1.34^{+1.96}_{-0.98}\)
    & 0.25 \\
    NFW & gOM & pHer & ALL
    & \(1.86^{+0.30}_{-0.26}\)
    & \(7.5^{+3.8}_{-3.0}\)
    & \(0.54^{+0.61}_{-0.45}\)
    & \(0.57^{+0.07}_{-0.09}\)
    & \(1.53^{+1.80}_{-1.02}\)
    & \(1.22^{+1.74}_{-0.87}\)
    & 0.51 \\
    \bottomrule
    \end{tabular}
    \label{tab:sims_red}
\end{table*}
\subsection{Gaseous tail}
\label{sec:Gaseous_tail}
\textsc{ORFEO} adopts a one-dimensional orbital-spine approximation in which the large-scale tail spine follows the reconstructed galaxy path, while ram-pressure drag causes stripped material to lag along it.
This approximation is motivated by hydrodynamical simulations of galaxies passing through cluster pericentre. Across the orbits and disc inclinations examined by
\citet{Roediger2007,Roediger2008}, the stripped gas forms wakes extending broadly along the galaxy's orbit, despite flaring and oscillations within the wake. These results motivate using the galaxy trajectory to describe the collective tail spine, rather than the detailed trajectories of individual gas elements. We therefore use the reconstructed trajectory from Sect.~\ref{sec:orbit_equations} to model the large-scale tail geometry and its LOS-velocity trend, without resolving the detailed filamentary or multiphase structure of the stripped
gas or the gas-removal mechanism itself. The scope of this approximation is discussed
in Appendix~\ref{app:tail_scope}.

We parametrise the reconstructed trajectory by its arc-length coordinate \(\gamma(t)\) so that \(\mathbf{r}(t)=\mathbf{r}[\gamma(t)]\). We discretise the stripped gas into parcels, each indexed by an assigned stripping time \(t_\mathrm{strip}\leq0\). A parcel initialised at \(t_\mathrm{strip}\) is placed at the galaxy position along the reconstructed trajectory and assigned the corresponding galaxy bulk orbital velocity along the trajectory; no local disc-rotation component is added,
\begin{equation}
\gamma_\mathrm{gas}(t_\mathrm{strip})=\gamma(t_\mathrm{strip}),
\qquad
\dot{\gamma}_\mathrm{gas}(t_\mathrm{strip})=\dot{\gamma}(t_\mathrm{strip}),
\label{eq:gas_initial_conditions}
\end{equation}
where dots denote time derivatives. The subsequent motion of the parcel along the tail spine is governed by
\begin{equation}
\ddot{\gamma}_\mathrm{gas}(t)
=
a_\parallel\!\left[\gamma_\mathrm{gas}(t)\right]
-
\frac{\dot{\gamma}_\mathrm{gas}(t)}
{\tau(t;t_\mathrm{strip})}.
\label{eq:gas_motion}
\end{equation}
The first term in Eq.~\eqref{eq:gas_motion} is the cluster gravitational acceleration projected along the reconstructed trajectory,
\begin{equation}
a_\parallel(\gamma)=
-\nabla\phi\!\left[\mathbf{r}(\gamma)\right]\cdot
\frac{\mathrm{d}\mathbf{r}}{\mathrm{d}\gamma},
\label{eq:aparallel}
\end{equation}
where \(\mathrm{d}\mathbf{r}/\mathrm{d}\gamma\) is the unit tangent because \(\gamma\) is an arc-length coordinate. The second term describes ram-pressure drag. For a parcel moving through the ICM, we write the characteristic drag timescale as \begin{equation}
\label{eq:tau}
\tau(t;t_\mathrm{strip}) =
\frac{2\,\Sigma_\mathrm{gas}(t;t_\mathrm{strip})}
{C_\mathrm{d}\,\rho_\mathrm{ICM}\!\left[\mathbf{r}_\mathrm{gas}(t)\right]\,
v_\mathrm{gas}(t)} ,
\end{equation}
with \(\mathbf{r}_\mathrm{gas}(t)=\mathbf{r}[\gamma_\mathrm{gas}(t)]\) and \(v_\mathrm{gas}(t)=|\dot{\gamma}_\mathrm{gas}(t)|\). Here \(\rho_\mathrm{ICM}\) is the local ICM mass density, \(C_\mathrm{d}=1\) is adopted as an effective order-unity drag coefficient, following commonly used astrophysical gas-drag prescriptions \citep{Roediger2007,Bekki2014}, while \(\Sigma_\mathrm{gas}(t;t_\mathrm{strip})\) is the effective gas surface density of the parcel. Since the drag acceleration depends on the ratio \(C_\mathrm{d}/\Sigma_\mathrm{gas}\), the normalisation of \(C_\mathrm{d}\) is degenerate with the adopted surface-density scale. The parcels are point-like Lagrangian tracers and no internal cloud geometry is assigned to them. Neglecting bulk motions of the ICM, \(v_\mathrm{gas}\) is the relevant parcel--ICM speed, so that the drag term in Eq.~\eqref{eq:gas_motion} corresponds to the standard ram-pressure acceleration \(C_\mathrm{d}\rho_\mathrm{ICM}v_\mathrm{gas}^2/(2\Sigma_\mathrm{gas})\), projected along the tail spine.

The effective surface density entering Eq.~\eqref{eq:tau}
parametrises differences in the net parcel--ICM coupling
associated with the gas region of origin and unresolved
post-stripping evolution. It is assigned at \(t_\mathrm{strip}\)
and held fixed during the subsequent parcel motion,
\(\Sigma_\mathrm{gas}(t;t_\mathrm{strip})
\equiv\Sigma_\mathrm{gas}(t_\mathrm{strip})\).
It should therefore be interpreted as an effective drag parameter
rather than as a resolved surface-density history of an individual
parcel. This effective coupling may also partly
represent unresolved release physics and the galaxy's gravitational attraction, which is not explicitly included in Eq.~\eqref{eq:gas_motion}, as discussed in Appendix~\ref{app:tail_scope}.
We anchor the assigned sequence to the observationally motivated surface-density interval introduced in Sect.~\ref{Sec:data_NGC}, bounded by the dense-disc reference value, \(\Sigma_\mathrm{disc,ref}\), and the adopted minimum effective surface density
for the diffuse outer-tail gas, \(\Sigma_\mathrm{tail,min}\). The assigned surface density approaches
\(\Sigma_\mathrm{disc,ref}\) for the most recently stripped parcels, while older parcels are assigned progressively lower values down to \(\Sigma_\mathrm{tail,min}\). For \(t_\mathrm{strip}<-t_\mathrm{scale}\), the assigned value is kept fixed at \(\Sigma_\mathrm{tail,min}\), avoiding an extrapolation below the adopted
surface-density floor.
Defining \(\Delta\Sigma\equiv\Sigma_\mathrm{disc,ref}-\Sigma_\mathrm{tail,min}\), we use
\begin{equation}
\label{eq:sigma_evol}
\Sigma_\mathrm{gas}(t_\mathrm{strip}) =
\begin{cases}
\Sigma_\mathrm{disc,ref}
+\dfrac{t_\mathrm{strip}}{t_\mathrm{scale}}\Delta\Sigma,
& t_\mathrm{strip}\in[-t_\mathrm{scale},0], \\[1.2ex]
\Sigma_\mathrm{tail,min},
& t_\mathrm{strip}\in[-t_\mathrm{max},-t_\mathrm{scale}).
\end{cases}
\end{equation}
Here \(t_\mathrm{max}\) is the maximum lookback time over which model parcels are sampled, chosen to be longer than the expected tail age. Equation~\eqref{eq:sigma_evol} is the simplest interpolation between the two observationally motivated surface-density anchors, introducing no additional shape parameter; it is not a resolved density-evolution equation for an individual parcel.
The parameter \(t_\mathrm{scale}\) in Eq.~\eqref{eq:sigma_evol} sets the timescale over which the assigned surface-density sequence spans the interval between the dense-disc and diluted-tail reference values. Since it is not known \emph{a priori}, we determine \(t_\mathrm{scale}\) through the scan described in Appendix~\ref{App:tscale_optimisation}, selecting the preferred fit among solutions for which \(t_\mathrm{scale}\) and \(T_\mathrm{tail}\) remain consistent within their uncertainties. The optimisation procedure, the validation of the effective surface-density prescription against hydrodynamical simulations, and the propagation of the associated uncertainty are described in Appendices~\ref{App:tail_validation} and \ref{App:tscale_optimisation}.

For each trial orbit, the integration of Eq.~\eqref{eq:gas_motion} gives the present-day position and velocity of all stripped parcels along the reconstructed trajectory. These quantities are mapped back to three-dimensional space through \(\mathbf{r}(\gamma)\) and projected onto sky position and LOS velocity. The resulting sequence defines the \textsc{ORFEO} model tail, which is then compared with the H\(\alpha\)-emitting regions described in Sect.~\ref{Sec:data_NGC}. The scope of the gaseous-tail approximation and the interpretation of the parcel stripping times are summarised in Appendix~\ref{app:tail_scope}.

\subsection{Data preparation and orbit fitting}
\label{sec:data_likelihood}

Equations~\eqref{eq:energy_conservation}--\eqref{eq:ang_evolution_unb} determine the galaxy trajectory once its present-day position and velocity in the cluster rest frame are specified. For a galaxy observed in projection, however, this state is only partially known: the measured quantities \((\mathrm{RA}_0,\mathrm{DEC}_0,cz_{\mathrm{obs},0})\) fix the sky position and observed recession velocity, but not the two Cartesian velocity components transverse to the LOS, \(v_{\perp,1}\) and \(v_{\perp,2}\). In addition, unless an independent estimate of the galaxy distance \(d_0\) is adopted, \(cz_{\mathrm{obs},0}\) cannot be decomposed uniquely into Hubble-flow and LOS peculiar-velocity contributions. \textsc{ORFEO} turns this incomplete information into a constrained inverse problem by exploring the missing components of the present-day state and selecting those for which the reconstructed orbit, combined with the gas-tail model, reproduces the observed spatial and kinematic spine of the stripped gas.

The observed position and recession velocity of NGC~4569 define the present-day boundary condition of the galaxy orbit, while the likelihood is evaluated from the tail measurements
\(\{\mathrm{RA}_i,\mathrm{DEC}_i,cz_{\mathrm{obs},i}\}_{i=1,\ldots,N}\),
described in Sect.~\ref{Sec:data_NGC}. The cluster centre, distance, LOS velocity, and gravitational potential described in Sects.~\ref{sec:Virgo} and~\ref{sec:Gravitational_potential} define the rest frame in which the orbit is integrated. For an assumed distance, the observed recession velocity is converted into the LOS peculiar velocity needed to initialise the orbit through
\begin{equation}
v_\parallel(d)
=
c\left[
\frac{1+cz_\mathrm{obs}/c}
{1+z_\mathrm{H}(d)}
-1
\right],
\qquad
z_\mathrm{H}(d)=\frac{H_0d}{c},
\label{eq:observed_red}
\end{equation}
where \(v_\parallel\) is measured along the LOS and \(c\) is the speed of light. In this work, \(d_0\) is treated as an external assumption rather than as a continuous fit parameter: we adopt \(d_0=d_\mathrm{cl}\) in the fiducial reconstruction and repeat the analysis over a discrete set of assumed distances in Sect.~\ref{sec:distance_scan}. For each fixed \(d_0\), the fitted parameter vector is therefore \(\Omega=(v_{\perp,1},v_{\perp,2})\). The LOS peculiar velocity is fixed by Eq.~\eqref{eq:observed_red} and included in the orbit integration. In the fixed Virgo-centred frame, \(v_{\perp,1}>0\) points west and \(v_{\perp,2}>0\) north. The \(v_{\rm RA}\) and \(v_{\rm DEC}\) quoted later describe the same reconstructed three-dimensional velocity projected onto the local east--north tangent plane at NGC~4569; the small numerical differences arise because this local tangent basis is not identical to the fixed Virgo-centred transverse basis.
The corresponding present-day position and velocity are then shifted to the cluster rest frame before orbit integration. Given a trial \(\Omega\), \textsc{ORFEO} reconstructs the galaxy orbit, evolves the stripped gas parcels as described in Sect.~\ref{sec:Gaseous_tail}, and projects the resulting model tail back onto the observable space \((\mathrm{RA},\mathrm{DEC},cz_\mathrm{obs})\). The projected model tail is then interpolated as a function of the gas-parcel time coordinate \(t\), giving the continuous model predictions \(\mathrm{RA}^\mathrm{m}(t;\Omega)\), \(\mathrm{DEC}^\mathrm{m}(t;\Omega)\), and \(cz^\mathrm{m}_\mathrm{obs}(t;\Omega)\). The model recession velocity is computed from the model distance and LOS peculiar velocity through Eq.~\eqref{eq:observed_red}. For each observed emitting region \(i\), we identify the corresponding point on the model tail by finding the value \(t_i^\star\) that minimises the dimensionless distance
\begin{eqnarray}
D_i^2(t;\Omega)=&&
\frac{\left[\mathrm{RA}_i-\mathrm{RA}^\mathrm{m}(t;\Omega)\right]^2}
{\sigma_{\mathrm{RA}}^2}
+
\frac{\left[\mathrm{DEC}_i-\mathrm{DEC}^\mathrm{m}(t;\Omega)\right]^2}
{\sigma_{\mathrm{DEC}}^2}
\nonumber \\
&+&
\frac{\left[cz_{\mathrm{obs},i}-cz^\mathrm{m}_{\mathrm{obs}}(t;\Omega)\right]^2}
{\sigma_{cz_{\mathrm{obs}}}^2},
\label{eq:model_data_association}
\end{eqnarray}
with \(t_i^\star=\arg\min_t D_i^2(t;\Omega)\). The agreement between the model tail and the data is then quantified through the \(\chi^2\)-like statistic
\begin{equation}
\chi^2_\mathrm{tail}(\Omega)=
\sum_{i=1}^{N} D_i^2(t_i^\star;\Omega)
=
\sum_{i=1}^{N}\min_t D_i^2(t;\Omega).
\label{eq:chi2}
\end{equation}
The quantities \(\sigma_{\mathrm{RA}}\), \(\sigma_{\mathrm{DEC}}\), and \(\sigma_{cz_{\mathrm{obs}}}\) are effective dispersions, not measurement errors on individual regions. This choice reflects the goal of the fit: the H\(\alpha\)-emitting regions trace an extended tail with finite width and internal velocity scatter, while the model describes its central spine. We therefore estimate these dispersions from the observed scatter of the tail around its mean spatial and kinematic trends, and use them both to associate observed regions with the model tail and to weight the residuals in Eq.~\eqref{eq:chi2}. With this definition, \(\chi^2_\mathrm{tail}\) measures how well a trial orbit reproduces the large-scale position and velocity structure of the stripped gas.

The best-fitting parameters, \(\Omega_\mathrm{BF}\), are obtained by minimising Eq.~\eqref{eq:chi2} with the Nelder--Mead algorithm \citep{NeldMead65}. This parameter set fixes the present-day cluster-frame state of the galaxy and therefore defines its reconstructed past orbit. We explore the solution space through \(\Delta\chi^2_\mathrm{tail}(\Omega)=\chi^2_\mathrm{tail}(\Omega)-\chi^2_{\mathrm{tail,min}}\) profiles and surfaces, which show how the reconstruction quality changes away from the best-fitting solution. The construction of the uncertainty budget from these profiles, and the propagation of the systematic contribution associated with the uncertainty on \(t_\mathrm{scale}\), are described in Appendix~\ref{App:tscale_optimisation}. A validation of the \(\chi^2\)-like reconstruction procedure on a controlled toy model is presented in Appendix~\ref{app:Likelihood}.

\subsection{Gravitational potential}
\label{sec:Gravitational_potential}

The orbit reconstruction requires specifying the gravitational field of the host cluster. In \textsc{ORFEO}, this field enters as an external spherically symmetric potential, \(\phi(r)\), which determines the acceleration acting on the galaxy and on the stripped gas parcels. This separates the orbit-fitting problem from the calibration of the cluster mass profile: the reconstruction can be run once a spherical potential has been specified from independent dynamical, X-ray, or lensing constraints. In the present application, we adopt a Navarro--Frenk--White (NFW) profile \citep{Navarro1996}, fully specified by the virial mass \(M_{200}\) and the scale radius \(R_\mathrm{s}\), calibrated for Virgo Cluster~A through the dynamical modelling described in Sect.~\ref{sec:manposst}.

For an NFW profile, the cumulative mass distribution is
\begin{equation}
\label{eq:Mass_NFW}
M_\mathrm{NFW}(r) =
M_{200}
\frac{
f(r/R_\mathrm{s})
}{
f(c_{200})
},
\qquad
f(x)=\ln(1+x)-\frac{x}{1+x},
\end{equation}
where \(c_{200}=r_{200}/R_\mathrm{s}\) is the concentration. The corresponding gravitational potential, defined to vanish at infinity, can be written as
\begin{equation}
\phi_\mathrm{NFW}(r)
=
-\frac{G M_{200}}{R_\mathrm{s} f(c_{200})}
\frac{\ln(1+r/R_\mathrm{s})}{r/R_\mathrm{s}},
\label{eq:Pot_NFW}
\end{equation}
where \(G\) is the gravitational constant. Equivalently, this expression corresponds to an NFW scale density \(\rho_\mathrm{s}\) defined by
\begin{equation}
4\pi R_\mathrm{s}^3 \rho_\mathrm{s}
=
\frac{M_{200}}{f(c_{200})}.
\label{eq:rho_s}
\end{equation}
The radius \(r_{200}\) is defined as the radius within which the mean enclosed density is \(200\) times the critical density of the Universe,
\begin{equation}
r_{200} =
\left(\frac{G M_{200}}{100 H_0^2}\right)^{1/3}.
\label{eq:r200}
\end{equation}
Once \(M_{200}\) and \(R_\mathrm{s}\) are specified, Eqs.~\eqref{eq:Mass_NFW}--\eqref{eq:r200} fully determine the cluster potential entering the orbit integration.

\subsubsection{Mass profile reconstruction with \textsc{MG-MAMPOSSt}}
\label{sec:manposst}

To calibrate the NFW potential defined above for Virgo Cluster~A, we infer \(M_{200}\) and \(R_\mathrm{s}\) from the projected phase-space distribution of cluster member galaxies. We use the \textsc{MG-MAMPOSSt} algorithm \citep{Pizzuti:2022ynt}, an extension of the \textsc{MAMPOSSt} method \citep{Mamon2013}, which constrains the mass profile, tracer-density profile, and velocity-anisotropy profile of a galaxy cluster under the assumptions of spherical symmetry and dynamical equilibrium.

Published estimates of the Virgo Cluster~A mass vary significantly, reflecting the different tracers and assumptions used to model the cluster, from X-ray hydrostatic analyses to galaxy kinematics \citep[e.g.][]{McLaughlin99,Conselice2001,Simionescu2017,Kashibadze2020}. Because \textsc{ORFEO} integrates galaxy orbits, we calibrate its gravitational potential directly from the projected phase-space distribution of cluster galaxies, which constrains the mass profile governing their motions, rather than adopting a single literature value. To assess the dependence on the tracer population, we perform the analysis for both the full member-galaxy sample and the early-type subsample described in Sect.~\ref{sec:Virgo}. The full analysis, reported in Appendix~\ref{app:manposst}, explores different combinations of mass model, projected tracer-density profile, and velocity-anisotropy prescription. Here we retain the NFW solutions, matching the mass profile adopted for the \textsc{ORFEO} potential. We adopt as fiducial the solution obtained from the early-type Virgo Cluster~A sample using an NFW mass profile, a generalised Osipkov--Merritt anisotropy profile, and a projected Hernquist tracer-density profile. The early-type population is expected to trace the virialised Virgo Cluster~A component more closely. The corresponding \(M_{200}\) and \(R_\mathrm{s}\), listed in Table~\ref{tab:sims_red}, define the fiducial potential, while the full-sample NFW solution is shown for comparison to illustrate the dependence on tracer selection.
\footnote{The reconstructed trajectory is largely insensitive to the adopted potential, while the bound status and long-term extrapolation are not. Using the lower-mass X-ray potential of \citet{Simionescu2017} yields orbital speeds and ram-pressure estimates consistent within \(1\sigma\) with the baseline reconstruction, but the orbit becomes formally unbound as the escape velocity decreases.}

\section{Results}
\label{Sec:results}
We use \textsc{ORFEO} to test whether the observed spatial and LOS-velocity structure of the H\(\alpha\) tail of NGC~4569 can be described by a coherent large-scale stripping sequence in the Virgo Cluster~A potential.
We first consider a fixed-distance reference case, adopting the canonical Virgo Cluster~A distance \(d_0=d_\mathrm{cl}=16.5\,\mathrm{Mpc}\) in Sect.~\ref{sec:fiducial_fixed_distance}, consistent with the observational analysis of NGC~4569 and its tail \citep{Sun2026}.
We then relax this assumption in Sect.~\ref{sec:distance_scan} and repeat the reconstruction over the LOS-distance range spanned by published estimates. Varying \(d_0\) changes both the three-dimensional position of NGC~4569 in the Virgo potential and the peculiar velocity inferred from its observed recession velocity, allowing us to quantify how the uncertain LOS position propagates into the reconstructed orbital phase and ram-pressure evolution.

\begin{figure}
\centering
\includegraphics[width=\columnwidth]{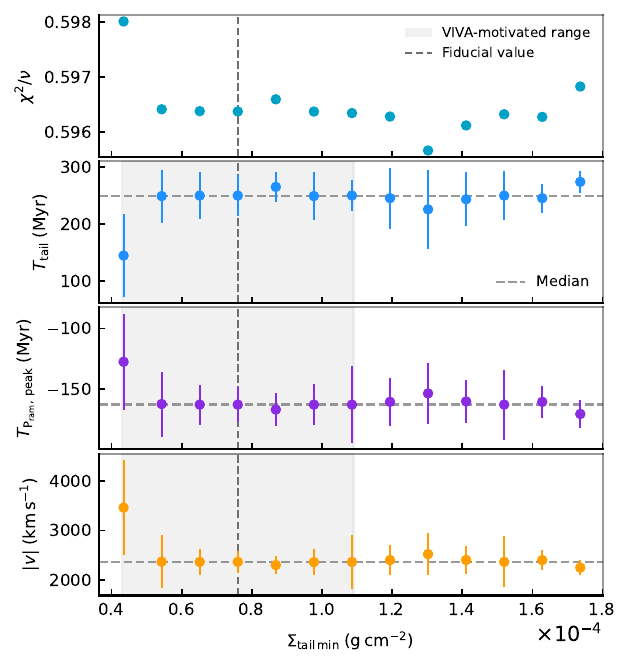}
\caption{
Sensitivity of reconstruction of NGC~4569 to the assumed minimum tail surface density, \(\Sigma_\mathrm{tail,min}\), for \(d_0=16.5\,\mathrm{Mpc}\).
From top to bottom, the panels show the reduced \(\chi^2\), the inferred tail age \(T_\mathrm{tail}\), the signed time of peak ram pressure \(t_{P_\mathrm{ram},\mathrm{peak}}\), and the present-day velocity amplitude \(|v|\).
Error bars indicate \(1\sigma\) uncertainties, while horizontal dashed lines show the median value across the explored grid.
The grey shaded region marks the range of \(\Sigma_\mathrm{tail,min}\) motivated by the VIVA H\,\textsc{i} data, and the vertical dashed line indicates the fiducial value adopted in this work.
}
\label{fig:NH_diagnostic_d016p5}
\end{figure}
\subsection{Reconstruction at the Virgo Cluster~A distance}
\label{sec:fiducial_fixed_distance}

We first test the sensitivity of the fixed-distance reconstruction to the adopted minimum tail surface density, \(\Sigma_\mathrm{tail,min}\), entering the tail parametrisation in Eq.~\eqref{eq:sigma_evol}. At \(d_0=d_\mathrm{cl}=16.5\,\mathrm{Mpc}\), we vary \(\Sigma_\mathrm{tail,min}\) over 13 values spanning \([4.3,\,17.3]\times10^{-5}\,\mathrm{g\,cm^{-2}}\), covering the observationally motivated range discussed in Sect.~\ref{Sec:data_NGC} and extending above it for robustness. For each value, we monitor the reduced goodness of fit, \(\chi^2/\nu\), with \(\nu=2N-2=186\); the tail age \(T_\mathrm{tail}\)\footnote{Throughout this work, \(T_{\rm tail}\) should therefore be interpreted as the ORFEO dynamical age of the fitted H\(\alpha\) phase-space sequence, not as the recombination or visibility time of the ionised gas, nor necessarily as the onset time of gas removal from NGC~4569.}; the signed time of maximum ram pressure, \(t_{P_\mathrm{ram},\mathrm{peak}}\); and the present-day velocity amplitude \(|v|\) in the cluster reference frame. Figure~\ref{fig:NH_diagnostic_d016p5} shows the dependence of these quantities on \(\Sigma_\mathrm{tail,min}\). Across the inspected grid, the reduced goodness of fit varies only at the few-\(10^{-3}\) level around \(\chi^2/\nu\simeq0.597\).
The inferred tail age, peak ram-pressure time, and present-day velocity amplitude remain consistent within their uncertainties and show no systematic trend, indicating that the fixed-distance reconstruction is only weakly sensitive to \(\Sigma_\mathrm{tail,min}\) over the explored range. We therefore adopt the central value of the observationally motivated interval, \(\Sigma_\mathrm{tail,min}^{\rm fid}=7.6\times10^{-5}\,\mathrm{g\,cm^{-2}}\); this choice is observationally motivated rather than selected by variations in the fit quality.

\begin{figure}
\centering
\includegraphics[width=\columnwidth]{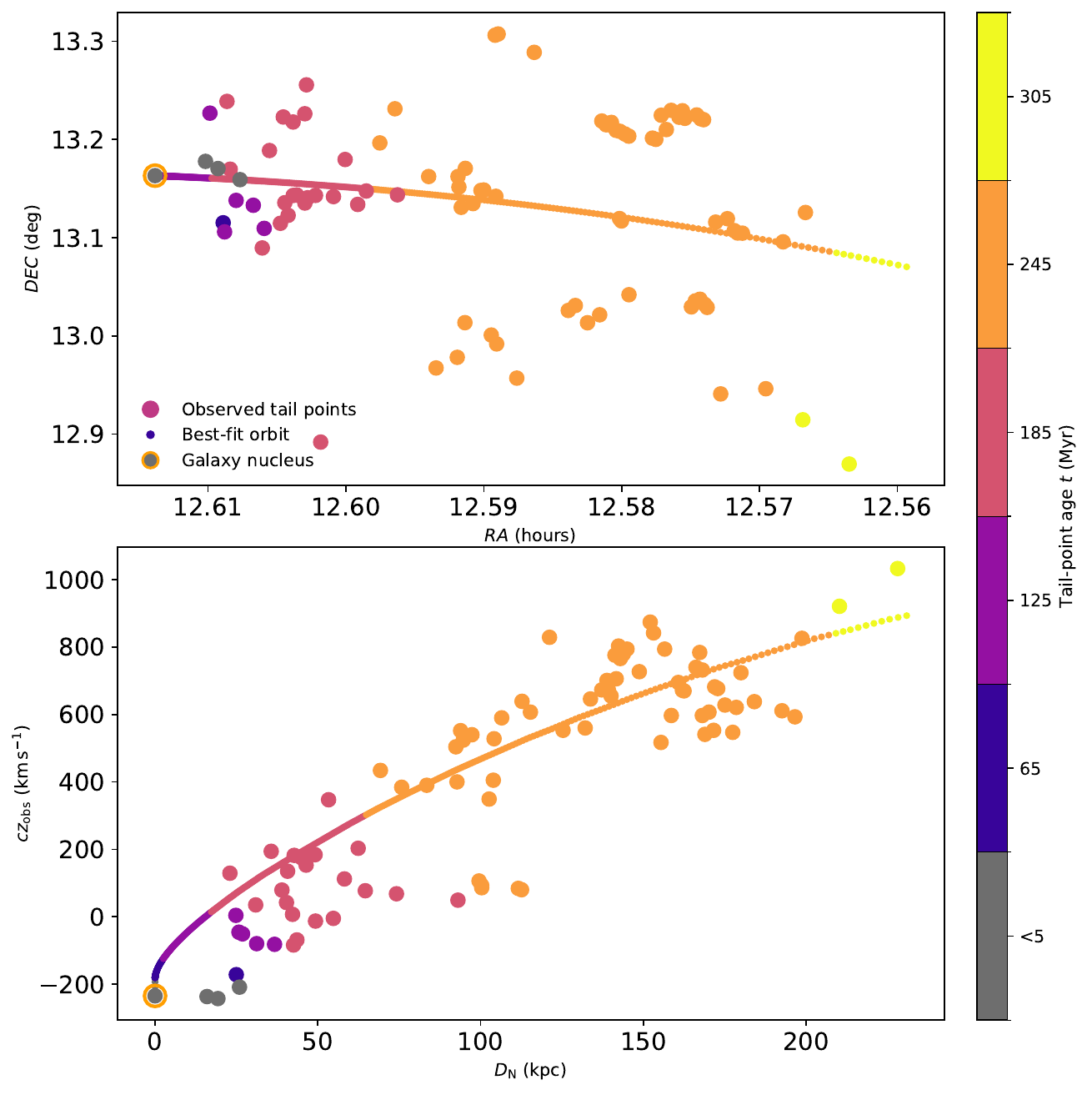}
\caption{
Phase-space reconstruction of NGC~4569 for \(d_0=16.5\,\mathrm{Mpc}\). The reconstructed tail from \textsc{ORFEO} is compared with the 94 H\(\alpha\)-emitting regions used in the fit, in the sky plane \((\mathrm{RA},\mathrm{DEC})\) and in the \(cz_\mathrm{obs}\)--\(D_\mathrm{N}\) projection. The colour coding indicates the model lookback time assigned by \textsc{ORFEO} along the reconstructed tail.}
\label{fig:Phase_space_orbit_d016p5}
\end{figure}
\subsubsection{Fiducial orbit}
\label{sec:fiducial_orbit_tail_chronology}

With the fiducial tail parametrisation fixed, we now reconstruct the dynamics of NGC~4569 at the canonical Virgo Cluster~A distance. Figure~\ref{fig:Phase_space_orbit_d016p5} shows the fiducial reconstruction at \(d_0=16.5\,\mathrm{Mpc}\). The reconstructed spine follows the projected orientation of the observed H\(\alpha\) tail in the sky plane and reproduces the monotonic increase of \(cz_\mathrm{obs}\) with projected distance from the galaxy in the \(cz_\mathrm{obs}\)--\(D_\mathrm{N}\) projection. This LOS phase-space agreement is the main constraint on the reconstruction: the fitted sequence must reproduce the kinematic ordering of the stripped gas, not only its projected spatial distribution. The scatter of the observed tracers around the reconstructed spine is expected because the model describes the large-scale tail as a one-dimensional sequence rather than resolving its width and internal structure.

The best-fitting parameters, listed in Table~\ref{tab:fiducial_orbit_ram_pressure}, place NGC~4569 on a high-velocity but energetically bound orbit in the adopted Virgo Cluster~A potential, with \(e=v^2/2+\phi<0\) and a velocity dominated by the transverse component \(v_{\perp,1}\). Along the reconstructed orbit, \textsc{ORFEO} assigns a model lookback time to each observed H\(\alpha\) region, shown by the colour coding in Fig.~\ref{fig:Phase_space_orbit_d016p5}. These times show a clear spatial ordering: tracers close to the disc form a compact group assigned to the most recent \(5\,\mathrm{Myr}\), \(t_{\rm strip}>-5\,\mathrm{Myr}\), whereas the extended tail follows the reconstructed spine and is assigned model lookback times older than \(\simeq75\,\mathrm{Myr}\). This apparent separation may partly reflect the spatial sampling of the observed H\(\alpha\) regions, but within the fitted sequence the extended tail is not associated with the most recent stripping times. The maximum lookback time assigned to the fitted H\(\alpha\) tracers gives \(T_\mathrm{tail}=250^{+37}_{-19}\,\mathrm{Myr}\). This age refers to the fitted ionised tail, not to the full duration of gas removal from NGC~4569. Older stripped material may exist beyond the currently detectable H\(\alpha\) phase and therefore does not enter the present reconstruction. The fiducial solution therefore interprets the extended H\(\alpha\) tail as a dynamically ordered structure sampling a substantial fraction of the recent reconstructed orbit rather than only its most recent phase.
\begin{figure}
\centering
\includegraphics[width=\columnwidth]{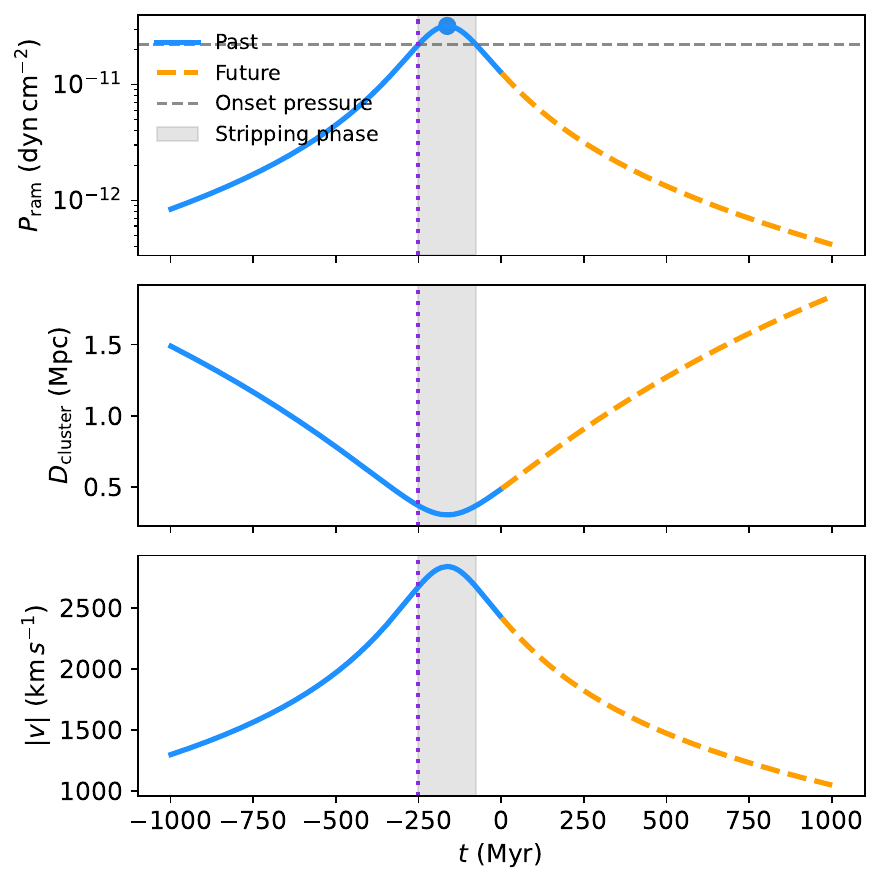}
\caption{
Ram-pressure history of NGC~4569 for the fiducial fixed-distance reconstruction.
The panels show, from top to bottom, the ram pressure \(P_\mathrm{ram}(t)\), the three-dimensional cluster-centric distance \(D_\mathrm{cluster}(t)\), and the velocity amplitude \(|v(t)|\) in the Virgo Cluster~A rest frame.
Time is defined such that \(t=0\) is the present epoch and \(t<0\) the past.
Solid blue curves show the reconstructed past evolution, while dashed orange curves show the forward integration from the present-day best-fitting state.
The blue marker indicates the maximum of \(P_\mathrm{ram}(t)\).
The horizontal dashed line marks \(P_\mathrm{ram}(-T_\mathrm{tail})\), the ram pressure at the lookback time of the oldest fitted H\(\alpha\) tail material.
The shaded region indicates the interval over which \(P_\mathrm{ram}(t)\geq P_\mathrm{ram}(-T_\mathrm{tail})\).
}
\label{fig:Pram_history_d016p5}
\end{figure}

\subsubsection{Ram-pressure history}
\label{sec:fiducial_ram_pressure_history}

We define
\(
P_\mathrm{ram}(t)=\rho_\mathrm{ICM}\!\left[D_\mathrm{cluster}(t)\right]\,v^2(t),
\)
where \(D_\mathrm{cluster}(t)\) is the reconstructed three-dimensional distance from the Virgo Cluster~A centre, \(v(t)\) is the velocity amplitude in the Virgo Cluster~A rest frame, and \(\rho_\mathrm{ICM}\) is evaluated from the Virgo gas-density model described in Sect.~\ref{sec:Virgo}. For \(t\leq0\), we use the reconstructed past trajectory; for \(t>0\), we use the forward integration from the best-fitting present-day state. Figure~\ref{fig:Pram_history_d016p5} shows that the fiducial reconstruction places NGC~4569 after the maximum of its ram-pressure event. The peak occurs at \(t_{P_\mathrm{ram},\mathrm{peak}}=-163^{+11}_{-15}\,\mathrm{Myr}\), near the closest approach to the Virgo Cluster~A centre, \(D_\mathrm{cluster}=305^{+19}_{-25}\,\mathrm{kpc}\), when the galaxy reaches \(|v|=2839^{+383}_{-191}\,\mathrm{km\,s^{-1}}\). Although this velocity exceeds the characteristic virial velocity of Virgo Cluster~A, the virial velocity is not an upper limit for an infalling galaxy, and the reconstructed orbit remains bound in the fiducial \textsc{MG-MAMPOSSt} potential. The high velocity is therefore a notable requirement of the fitted tail phase-space structure.

Taking the ram pressure at the model lookback time of the oldest fitted tail tracer, \(P_\mathrm{ram}(-T_\mathrm{tail})\), as a reference, we find
\(
P_\mathrm{ram,peak} > P_\mathrm{ram}(-T_\mathrm{tail}) > P_\mathrm{ram}(0).
\)
Thus, the present-day ram pressure is lower than that along the reconstructed orbit at the age of the oldest fitted H\(\alpha\) tracer. Since stripping is expected to proceed broadly from the outer towards the inner disc, where the gas surface density and restoring force are higher, continued removal of the remaining gas would generally require a ram pressure at least comparable to, and likely higher than, the level associated with earlier stripping phases. The ordering \(P_\mathrm{ram}(-T_\mathrm{tail})>
P_\mathrm{ram}(0)\), together with the expected stronger binding of the retained disc gas, therefore suggests that the main disc-stripping episode has ended by the present epoch. In this interpretation, stripping continues after the ram-pressure maximum until the declining pressure becomes insufficient to remove the remaining disc gas, while more diffuse material may still be removed. The last post-peak crossing of this reference level occurs at \(t_\mathrm{end}\simeq-75\,\mathrm{Myr}\), approximately coinciding with the youngest model ages assigned to the extended tail in Sect.~\ref{sec:fiducial_orbit_tail_chronology}. We adopt this crossing as an indicative end of the main disc-stripping episode. A quantitative test of this interpretation using the restoring force at the observed disc truncation radius is presented in Sect.~\ref{sec:Physical_solutions}.

\begin{table}
\centering
\caption{
Fiducial fixed-distance reconstruction quantities for \(\Sigma_\mathrm{tail,min}^{\rm fid}=7.6\times10^{-5}\,\mathrm{g\,cm^{-2}}\).
Uncertainties are total uncertainties, including the contribution associated with the propagated uncertainty on \(t_\mathrm{scale}\).
The value \(P_\mathrm{ram}(-T_\mathrm{tail})\) is evaluated at the model lookback time of the oldest fitted H\(\alpha\) tracer. The time \(t_\mathrm{end}\) denotes the post-peak epoch
at which \(P_\mathrm{ram}\) drops back to
\(P_\mathrm{ram}(-T_\mathrm{tail})\), used as an indicative end of the main
disc-stripping episode}.

\label{tab:fiducial_orbit_ram_pressure}
\begin{tabular}{lc}
\hline
Quantity & Value \\
\hline
\multicolumn{2}{c}{Orbit and tail quantities} \\
\hline
\(v_{\perp,1}\) & \(-1876^{+275}_{-219}\,\mathrm{km\,s^{-1}}\) \\
\(v_{\perp,2}\) & \(61^{+80}_{-83}\,\mathrm{km\,s^{-1}}\) \\
\(T_\mathrm{tail}\) & \(250^{+37}_{-19}\,\mathrm{Myr}\) \\
\hline
\multicolumn{2}{c}{Ram-pressure quantities} \\
\hline
\(t_{P_\mathrm{ram},\mathrm{peak}}\) & \(-163^{+11}_{-15}\,\mathrm{Myr}\) \\
\(P_\mathrm{ram,peak}\) & \(\left(3.20^{+0.68}_{-0.73}\right)\times10^{-11}\,\mathrm{dyn\,cm^{-2}}\) \\
\(P_\mathrm{ram}(-T_\mathrm{tail})\) & \(\left(2.20^{+0.57}_{-0.41}\right)\times10^{-11}\,\mathrm{dyn\,cm^{-2}}\) \\
\(P_\mathrm{ram}(0)\) & \(\left(1.27^{+0.21}_{-0.19}\right)\times10^{-11}\,\mathrm{dyn\,cm^{-2}}\) \\
\(t_\mathrm{end}\) & \(\simeq -75\,\mathrm{Myr}\) \\
\hline
\end{tabular}
\end{table}

\subsection{Exploring distance assumptions}
\label{sec:distance_scan}
Recent surface-brightness-fluctuation (SBF) measurements by \cite{Cantiello24} place NGC~4569 at larger distances than the canonical Virgo Cluster~A value, with the catalogue reference estimate \(d_\mathrm{ref}=17.47\pm1.06\,\mathrm{Mpc}\), derived using their preferred SBF--colour calibration, and \(d_\mathrm{mean}=18.09\pm1.23\,\mathrm{Mpc}\), obtained by averaging the estimates from all available calibrations. To assess how different assumed distances change the inferred orbit and ram-pressure history of NGC~4569, we therefore repeat the full reconstruction over an interval encompassing both \(d_\mathrm{cl}=16.5\,\mathrm{Mpc}\) and the SBF-based estimates.

\begin{figure}
\centering
\includegraphics[width=\columnwidth]{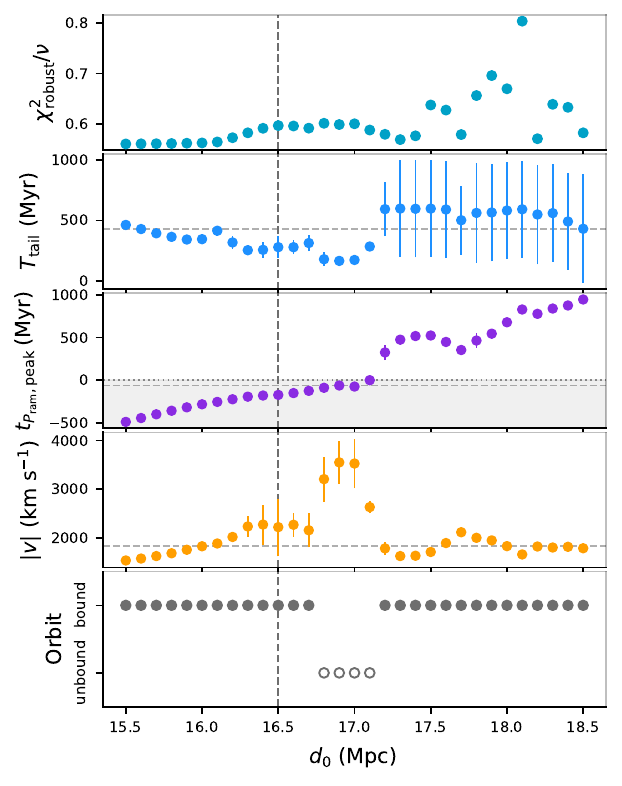}
\caption{
Sensitivity of the \textsc{ORFEO} reconstruction to the assumed LOS distance of NGC~4569. From top to bottom: reduced robust goodness of fit, tail age, time of peak ram pressure, orbital velocity amplitude, and orbital state. The vertical dashed line marks the fiducial Virgo Cluster~A distance. 
}
\label{fig:distance_scan_continuity}
\end{figure}

Figure~\ref{fig:distance_scan_continuity} shows the reconstruction for 36 fixed distances spanning \(d_0\in[15.0,18.5]\,\mathrm{Mpc}\) in steps of \(0.1\,\mathrm{Mpc}\), refitting \(v_{\perp,1}\) and \(v_{\perp,2}\) at each value. The reduced robust goodness of fit remains close to \(\chi^2_\mathrm{robust}/\nu\simeq0.54\)--\(0.64\) over most of the scan, showing that the projected spatial and LOS-velocity structure of the H\(\alpha\) tail can be reproduced over a broad range of assumed distances. The fit quality alone therefore does not select a unique distance for NGC~4569. The first clear change appears in the timing of the ram-pressure maximum. At low assumed distances, \(t_{P_\mathrm{ram},\mathrm{peak}}<0\), so NGC~4569 is at present after the maximum of its ram-pressure event, as in the fiducial solution. As \(d_0\) increases, \(t_{P_\mathrm{ram},\mathrm{peak}}\)
rises steadily, crosses the present epoch around
\(17.1\,\mathrm{Mpc}\), and becomes positive at larger
distances. This crossing corresponds to a change in the present-day cluster-centric radial velocity from outward to inward motion, placing the galaxy on the post-pericentre and pre-pericentre branches, respectively
(Appendix~\ref{app:orbital_diagnostics}). In the high-distance solutions, the galaxy is therefore reconstructed before peak ram pressure, with the peak shifted up to \(\sim0.8\)--\(1.0\,\mathrm{Gyr}\) into the future near the upper end of the explored range. The same distance interval also contains a narrow window of unbound or marginally bound solutions. This identifies \(d_0\simeq16.8\)--\(17.1\,\mathrm{Mpc}\) as a transition interval in the reconstructed orbital phase, rather than as a sharply measured critical distance.
The tail age and velocity amplitude support the same transition but are less regular. Along the low-distance branch, \(T_\mathrm{tail}\) generally decreases from \(\simeq0.46\,\mathrm{Gyr}\) at \(15.5\,\mathrm{Mpc}\) to \(\simeq{\rm 0.25}\,\mathrm{Gyr}\) at the fiducial distance. Across \(d_0\simeq16.8\)--\(17.1\,\mathrm{Mpc}\), it reaches \(\simeq0.17\)--\(0.28\,\mathrm{Gyr}\) before increasing to \(\simeq0.6\,\mathrm{Gyr}\) on the high-distance branch, where inferred ages are typically \(0.4\)--\(0.6\,\mathrm{Gyr}\) with larger uncertainties. The reconstructed \(|v|\) likewise increases along the low-distance branch, becomes highly structured across the transition, and settles onto a lower-amplitude high-distance branch. The sign change of \(t_{P_\mathrm{ram},\mathrm{peak}}\) therefore remains the clearest indicator of the transition, while \(T_\mathrm{tail}\) and \(|v|\) near and beyond it should be interpreted more cautiously. The reconstructed eccentricities provide further context for the plausibility of these orbital families through comparison with recently infalling satellites in cosmological simulations \citep{Gill2004}; the comparison is presented in Appendix~\ref{app:orbital_diagnostics}.

\begin{figure}
\centering
\includegraphics[width=\columnwidth]{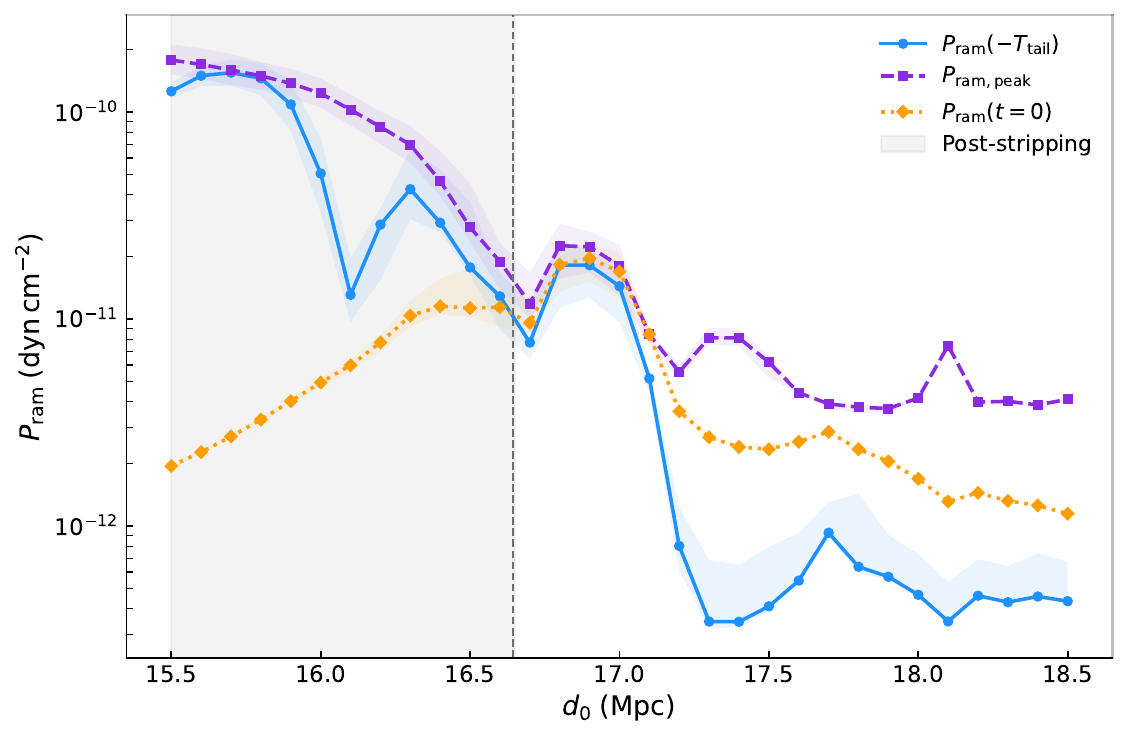}
\caption{
Ram-pressure regimes as a function of the assumed LOS distance of NGC~4569. The curves show \(P_\mathrm{ram}(-T_\mathrm{tail})\), \(P_\mathrm{ram,peak}\), and \(P_\mathrm{ram}(0)\). The grey region marks
\(P_\mathrm{ram}(-T_\mathrm{tail})>P_\mathrm{ram}(0)\).
Together with the expected stronger binding of the retained disc gas, this ordering supports the qualitative interpretation that the main disc-stripping episode has ended (Sect.~\ref{sec:fiducial_ram_pressure_history}).}

\label{fig:distance_scan_pram}
\end{figure}

The ram-pressure diagnostics in Fig.~\ref{fig:distance_scan_pram}
show how the conditions for disc stripping vary across the reconstructed orbital families. As in Sect.~\ref{sec:fiducial_ram_pressure_history}, the ordering \(P_\mathrm{ram}(-T_\mathrm{tail})>
P_\mathrm{ram}(0)\), combined with the expected stronger binding of the retained gas, qualitatively favours a
configuration in which the main disc stripping episode has ended. This condition is satisfied on the low-distance branch up to \(d_0\simeq16.6\,\mathrm{Mpc}\), where the reconstructed peak pressure is of order \(10^{-10}\,\mathrm{dyn\,cm^{-2}}\) and exceeds the present-day value. Across \(d_0\simeq16.6\)--\(17.1\,\mathrm{Mpc}\), \(P_\mathrm{ram,peak}\), \(P_\mathrm{ram}(-T_\mathrm{tail})\), and \(P_\mathrm{ram}(0)\) become comparable, coinciding with the orbital-phase transition identified above. At larger distances, \(P_\mathrm{ram}(-T_\mathrm{tail})\) drops by about an order of magnitude below \(P_\mathrm{ram}(0)\), while the ram-pressure maximum shifts to the future. The pressure ordering thus complements the peak-time diagnostic by describing the evolution of the stripping conditions. Near the transition, the present pressure can remain above \(P_\mathrm{ram}( -T_\mathrm{tail})\) even after the ram-pressure maximum has passed.

The distance scan therefore yields two evolutionary families rather than a unique distance for NGC~4569, with the central SBF estimates of \cite{Cantiello24} lying on the high-distance, pre-peak side of the transition. The existence of this change in evolutionary interpretation is robust, whereas its exact boundary and the detailed properties of the high-distance branch should not be over-interpreted because the reconstructed \(T_\mathrm{tail}\), \(|v|\), and pressure diagnostics become increasingly structured and branch-dependent across the transition. In Sect.~\ref{sec:Physical_solutions}, we test the physical plausibility of these solutions against an independent stripping threshold for NGC~4569.

\section{Discussion}\label{sec:discussion}
To interpret the reconstructed ram-pressure histories in terms of gas removal from NGC~4569, we compare them with the pressure required to strip gas at the observed disc truncation radius, estimated independently from the galaxy structural and gas properties.
\subsection{Physical plausibility of the distance-dependent solutions}
\label{sec:Physical_solutions}

The distance scan in Sect.~\ref{sec:distance_scan} yields two families of formally acceptable reconstructions: a low-distance post-peak branch and a high-distance branch in which the ram-pressure maximum lies in the future. The effective pressure required for gas removal depends on both the wind--disc geometry and the gas distribution at the stripping epoch \citep[e.g.][]{RoedigerBruggen2006}. A fully orbit-dependent threshold would therefore require combining the reconstructed wind--disc angle (Appendix~\ref{app:wind_disc_angle}) with an evolving gas-disc model and a geometry-dependent stripping efficiency. For a uniform comparison among the reconstructed orbits, we instead adopt a common physical reference: the pressure required to strip gas at the observed disc truncation radius of NGC~4569.

To estimate this pressure scale, we adopt a Gunn--Gott-type restoring-force criterion \citep{Gunn72}. Stripping at a radius \(R_\mathrm{strip}\) is expected when the external ram pressure exceeds the maximum vertical restoring force per unit area acting on the gas,
\begin{equation}
P_{\rm ram,th}(R_{\rm strip}) =
\Sigma_{\rm gas}(R_{\rm strip})\,
\max_{z}\left|g_z(R_{\rm strip},z)\right| .
\label{eq:Pram_threshold}
\end{equation}
Here, \(g_z(R_{\rm strip},z)\) is the vertical gravitational acceleration generated by the adopted disc mass model. We compute \(g_z\) with \texttt{galpy} \citep{Bovy2015}. 
We model the stellar, atomic, and molecular discs with exponential surface-density profiles,
\begin{equation}
\Sigma_i(R)=\Sigma_{i,0}\exp(-R/r_{h,i}),
\end{equation}
where \(i=\star,\HI{},\Htwo\), and \(r_{h,i}\) is the corresponding scale radius. For an exponential disc normalised to a total mass \(M_i\) and truncated at \(R_{\max}\), the central surface density is
\begin{equation}
\Sigma_{i,0} =
\frac{M_i}{2\pi r_{h,i}\left[r_{h,i}-e^{-R_{\max}/r_{h,i}}(R_{\max}+r_{h,i})\right]} .
\end{equation}
In practice, \(R_{\max}\) is chosen sufficiently large that the normalisation is effectively equivalent to integrating the exponential profile to infinity.

For NGC~4569, we evaluate Eq.~\eqref{eq:Pram_threshold} at the observed truncation radius, \(R_\mathrm{strip}=4.8\,\mathrm{kpc}\). We adopt a stellar scale radius
\(r_h=0.59R_e=3.7\,\mathrm{kpc}\), using
\(R_e=6.26\,\mathrm{kpc}\) from \citet{Munoz-Mateos2015}, and a stellar mass \(M_\star=10^{10.74}\,M_\odot\) \citep{Boselli23a}. The pre-stripping atomic gas mass is estimated from the H\,\textsc{i} deficiency: we adopt the observed mass \(M_{\mathrm{H\,I}}=10^{8.90}\,M_\odot\) and deficiency \(\mathrm{H\,I\!-\!def}=1.08\) from GOLDMine \citep{Gavazzi03}, updated to \(d_0=16.5\,\mathrm{Mpc}\).
Together, these give a pre-stripping mass
\(M_{\mathrm{H\,I}}\simeq10^{9.98}\,M_\odot\). Motivated by the systematically larger radial extent of \HI{} discs relative to stellar discs in nearby spiral galaxies \citep[e.g.][]{BroeilsRhee1997}, we adopt an \HI{} disc scale radius \(1.8\) times larger than the stellar one. We estimate the pre-stripping molecular gas mass as \(M_{\mathrm{H}_2}\simeq10^{10.0}\,M_\odot\), corresponding to \(M_{\mathrm{H}_2}/M_{\mathrm{H\,I}}\simeq1\), using the scaling relations of \citet{Boselli14a}, and assume that this component follows the stellar scale radius. The truncated CO distribution suggests that the interaction may also have removed molecular gas
\citep{Boselli06,Boselli14b}, so the present-day mass
need not represent the pre-stripping reservoir. The lower VERTICO estimate,
\(M_{\mathrm{H}_2}\simeq10^{9.58}\,M_\odot\) \citep{Brown2021}, is consistent with this interpretation. 

We propagate the uncertainties with a Monte Carlo procedure, assuming \(0.15\) dex uncertainty on \(M_\star+M_{\Htwo}\), \(0.3\) dex on \(M_{\HI}\), motivated by the scatter of the \HI{}-deficiency calibration, and \(20\,\%\) uncertainty on the adopted scale radii. The resulting threshold distribution has
a median and 16th--84th percentile interval of
\begin{equation}
P_{\rm ram,th}(R_\mathrm{strip}=4.8\,\mathrm{kpc})
=
6.14^{+10.05}_{-3.44}\times10^{-11}\,
\mathrm{dyn\,cm^{-2}} .
\label{eq:Pram_threshold_value}
\end{equation}
This value provides an external physical scale against which the \textsc{ORFEO} ram-pressure histories can be tested. The threshold uncertainty reflects uncertainties
in the stellar mass, pre-stripping gas content, and disc scale
radii, which determine the local restoring pressure, and is
conditional on the adopted disc profiles and stripping criterion. The reconstructed pressure uncertainties propagate
those in the fitted velocity components and \(t_\mathrm{scale}\); additional systematic uncertainties associated with the cluster potential, ICM density profile, and effective tail
prescription are not included.

\begin{figure}
\centering
\includegraphics[width=\columnwidth]{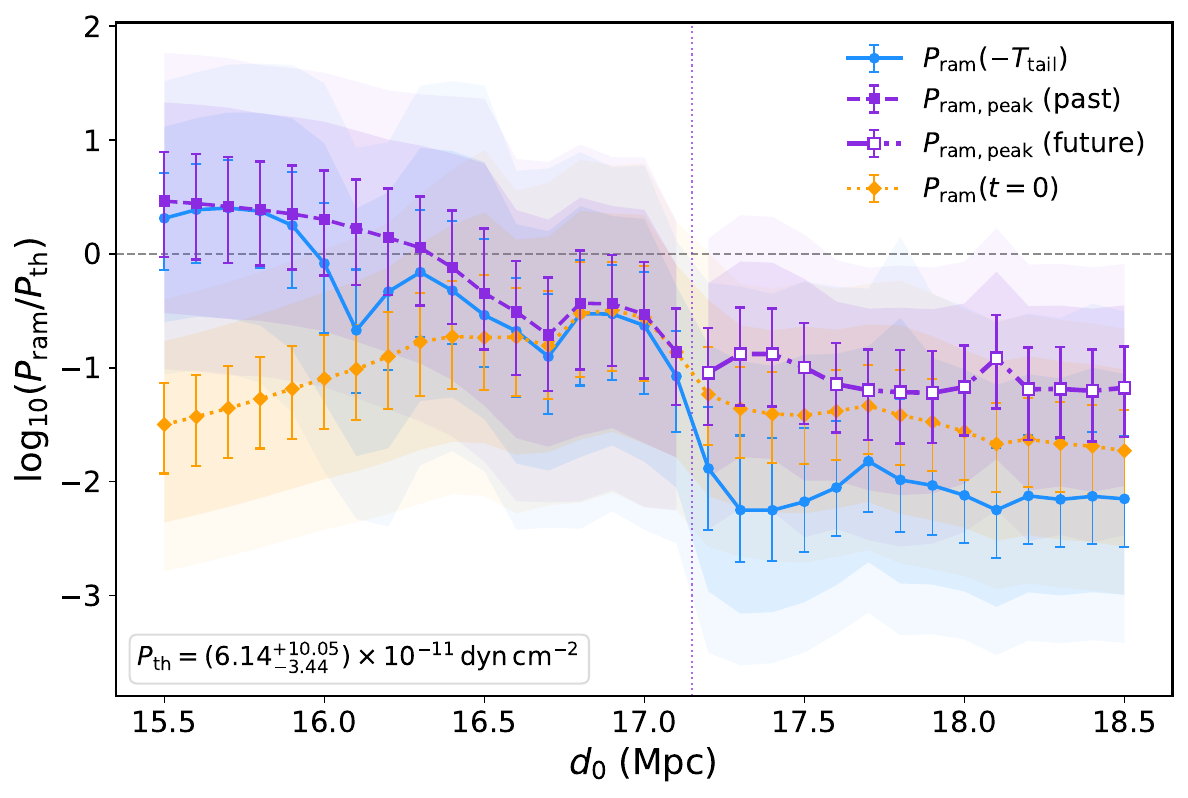}
\caption{
Comparison between the reconstructed ram-pressure histories and the stripping threshold at the observed disc truncation radius of NGC~4569. The curves show \(\log_{10}(P_i/P_{\rm ram,th})\) for \(P_{\rm ram}(-T_{\rm tail})\), \(P_{\rm ram,peak}\), and \(P_{\rm ram}(t=0)\) as a function of LOS distance. The dashed horizontal line marks equality with \(P_{\rm ram,th}=6.14^{+10.05}_{-3.44}\times10^{-11}\,\mathrm{dyn\,cm^{-2}}\); error bars and shaded regions show the combined \(1\sigma\) and 2--\(3\sigma\) uncertainties, respectively. For orbits whose ram-pressure maximum occurs
in the future (\(d_0\geq17.2\,\mathrm{Mpc}\)), peaks are shown
with open violet squares; past peaks use filled squares.
The vertical dotted line separates these regimes.}

\label{fig:pram_threshold_residuals}
\end{figure}
Figure~\ref{fig:pram_threshold_residuals} compares the stripping threshold with three characteristic pressure levels from the distance-dependent reconstructions: \(P_\mathrm{ram,peak}\), \(P_\mathrm{ram}(0)\), and \(P_\mathrm{ram}(-T_\mathrm{tail})\). These probe complementary aspects of the inferred history. \(P_\mathrm{ram,peak}\) tests whether the reconstructed orbit reaches conditions sufficient to produce the observed disc truncation, \(P_\mathrm{ram}(0)\) whether such conditions are present today, and \(P_\mathrm{ram}(-T_\mathrm{tail})\) whether they were already reached at the model lookback time of the oldest fitted H\(\alpha\) tracer. We quantify the comparison through \(\log_{10}[P_i(d_0)/P_{\rm ram,th}]\), where \(P_i=\{P_\mathrm{ram,peak},P_\mathrm{ram}(0),P_\mathrm{ram}(-T_\mathrm{tail})\}\); positive values indicate pressures above the adopted stripping threshold. The central pressure comparison favours somewhat lower distances than the fiducial value.
The propagated uncertainties and the dependence on the adopted disc and cluster models nevertheless limit the precision of a distance constraint based on this diagnostic.
\begin{table}
\centering
\caption{
Maximum sampled distance at which each reconstructed ram-pressure diagnostic remains compatible with the stripping threshold at the observed disc truncation radius of NGC~4569.
}
\label{tab:pram_threshold_compatibility}
\begin{tabular}{lcc}
\hline
Pressure diagnostic & \(1\sigma\) & \(3\sigma\) \\
                    & \(d_{0,\max}\,[\mathrm{Mpc}]\) & \(d_{0,\max}\,[\mathrm{Mpc}]\) \\
\hline
\(P_\mathrm{ram,peak}\) & 16.8 & 18.1 \\
\(P_\mathrm{ram}(-T_\mathrm{tail})\) & 16.5 & 17.8 \\
\(P_\mathrm{ram}(0)\) & -- & 17.2 \\
\hline
\end{tabular}
\tablefoot{
The limits are computed from the sampled distance grid shown in Fig.~\ref{fig:pram_threshold_residuals}. The dash indicates that no sampled distance is compatible at the stated uncertainty level.
}
\end{table}

Table~\ref{tab:pram_threshold_compatibility} summarises the compatibility limits. At the \(1\sigma\) level, only reconstructed past pressures on the low-distance branch are compatible with the stripping threshold: \(P_\mathrm{ram,peak}\) remains compatible up to \(d_0=16.8\,\mathrm{Mpc}\) and \(P_\mathrm{ram}(-T_\mathrm{tail})\) up to \(d_0=16.5\,\mathrm{Mpc}\), whereas \(P_\mathrm{ram}(0)\) remains below the threshold throughout the scan. Thus, within the nominal uncertainty budget, the present-day interaction is not sufficient to strip gas from the remaining disc at its observed truncation radius, whereas the reconstructed past event is compatible with such stripping. This comparison is conservative for the onset of stripping, since the gaseous disc was likely more extended at earlier times and its outer gas would therefore have required a lower ram pressure to be removed. The low-distance solutions therefore support a post-stripping interpretation and imply that the fitted H\(\alpha\) tail may trace a later observable phase of an episode that began earlier. 

At the \(3\sigma\) level the comparison is less restrictive, but the physical preference for the low-distance branch remains. \(P_\mathrm{ram,peak}\) is compatible with the stripping threshold up to \(d_0\simeq18.1\,\mathrm{Mpc}\); however, for \(d_0\gtrsim17.1\,\mathrm{Mpc}\) this peak occurs in the future and therefore cannot be associated with the event responsible for the observed disc truncation. Independent constraints from the truncated young stellar populations place the associated quenching/stripping episode at lookback times of order \(10^8\,\mathrm{yr}\) \citep[e.g.][]{Boselli06,Crowl08,Boselli16a}. In the same high-distance regime, neither \(P_\mathrm{ram}(-T_\mathrm{tail})\) nor \(P_\mathrm{ram}(0)\) is robustly compatible with the threshold, apart from small branch-sensitive oscillations. The high-distance histories therefore do not naturally provide the pressure required at an epoch consistent with the independently inferred truncation time.

The threshold comparison also clarifies the physical implications of the three distance anchors considered here. At the fiducial Virgo Cluster~A distance, \(d_0=16.5\,\mathrm{Mpc}\), the reconstructed past ram-pressure event is compatible with stripping at the observed disc radius, whereas \(P_\mathrm{ram}(0)\) remains below the threshold. The SBF-based distances from \cite{Cantiello24} are less favoured by this physical consistency test: \(d_\mathrm{ref}=17.47\,\mathrm{Mpc}\) is compatible only within the broader \(3\sigma\) envelope, while \(d_\mathrm{mean}=18.09\,\mathrm{Mpc}\) lies at the upper edge of the \(3\sigma\) peak-pressure compatibility range. At both distances, however, the reconstructed ram-pressure maximum occurs in the future and therefore cannot explain the past disc truncation. High-distance solutions can therefore reproduce the tail phase space but do not naturally explain the disc truncation and H\(\alpha\) tail at an epoch consistent with the independent stellar-population constraints.
Independent constraints do not fully resolve this distance ambiguity. \citet{Boselli16a}, referring to the TRGB-based Virgo infall analysis of \citet{Karachentsev14}, noted that NGC~4569 may lie at \(D\gtrsim17\,\mathrm{Mpc}\), close to or beyond the transition between the reconstructed branches. Conversely, \citet{koppen18} classified NGC~4569 as a past stripper from the surviving \HI{} disc and the inferred past-to-present ram-pressure contrast, independently supporting the post-stripping interpretation of the low-distance branch. Thus, the available distance estimates alone do not uniquely select between the reconstructed families.

\subsection{Comparison with previous models of NGC~4569}
\label{sec:comparison_previous_models}

NGC~4569 has long been interpreted as one of the clearest Virgo examples of advanced ram-pressure stripping. A key timing constraint was obtained by \citet{Boselli06}, who modelled the multi-wavelength radial profiles of NGC~4569, including the gas, H\(\alpha\), UV, optical, and near-infrared emission. They showed that starvation alone cannot reproduce the observed truncation of the gas and star-forming discs, whereas ram-pressure stripping can account for the radial quenching pattern. Their best-fitting model dates the peak of the stripping event to \(\simeq100\,\mathrm{Myr}\) ago. The comparatively regular stellar disc, despite the strongly disturbed gaseous component, favours a predominantly hydrodynamical origin for the tail, although an additional gravitational contribution from an interaction with another galaxy or with the cluster potential cannot be excluded.

The low-distance branch of \textsc{ORFEO} is consistent with this independent constraint. For \(d_0\leq16.5\,\mathrm{Mpc}\), the reconstruction places NGC~4569 after the maximum of its ram-pressure history, with \(t_{P_{\rm ram},\mathrm{peak}}\leq-163\,\mathrm{Myr}\). Combined with the threshold analysis, which indicates that the present-day ram pressure is insufficient to strip gas from the remaining disc at its observed truncation radius, this supports interpreting the low-distance solutions as both post-peak and post-stripping. The Boselli et al. and \textsc{ORFEO} timescales are not identical measurements: the former is inferred from radial truncation and stellar-population gradients in the disc, whereas the latter describes the orbital ram-pressure history and the chronology of the observed ionised tail. A complementary constraint is provided by \citet{Sun2026}, who estimated that gas along the extended H\(\alpha\) tail may require \(\sim500\,\mathrm{Myr}\) to decelerate towards the Virgo systemic velocity. This timescale is more directly comparable with \(T_\mathrm{tail}\) than with \(t_{P_{\rm ram},\mathrm{peak}}\), although the estimates rely on different dynamical assumptions.

This interpretation is also broadly consistent with the \HI{} modelling of \citet{Vollmer04}, based on 21-cm observations of the atomic gas distribution and kinematics. Their \(N\)-body/sticky-particle simulations favoured a post-peak scenario over ongoing stripping, with the imposed ram-pressure pulse peaking a few \(10^2\,\mathrm{Myr}\) before the best-matching snapshot. We therefore use the Vollmer et al. model as independent support for the post-stripping state, while the Boselli et al. radial-profile and stellar-population analysis provides the more direct comparison for the timing of disc truncation.

The simulation approach of \citet{Vollmer04} and the reconstruction performed here provide complementary information. Tailored simulations follow the hydrodynamical response of the gas disc and stripped material to an assumed ram-pressure history. \textsc{ORFEO} does not model this hydrodynamical response; instead, it uses the observed phase-space structure of the tail to reconstruct the galaxy orbit, tail chronology, and ram-pressure history. In this sense, tailored simulations predict the gas response to a prescribed interaction history, whereas \textsc{ORFEO} infers the recent dynamical history directly from the observed tail and can provide constrained input for such simulations.

The H\(\alpha\) morphology of \citet{Boselli16a} also suggested a geometrical interpretation of the interaction. They argued for a projected west-to-east orbit and proposed that NGC~4569 entered the cluster from behind, encountering the dense Virgo environment north of M87 at a cluster-centric distance of \(\gtrsim230\,\mathrm{kpc}\). For \(d_0=16.5\,\mathrm{Mpc}\), the \textsc{ORFEO} reconstruction gives \(v_\mathrm{RA}=+1915\,\mathrm{km\,s^{-1}}\) and \(v_\mathrm{DEC}=+73\,\mathrm{km\,s^{-1}}\) after projection of the reconstructed three-dimensional velocity onto the local east--north tangent plane at NGC~4569. The projected motion is therefore dominated by the RA component, corresponding to a mostly west-to-east trajectory with only a small south-to-north component; this remains true across the low-distance branch. The reconstructed closest approach likewise remains within a few \(10^2\,\mathrm{kpc}\), with \(D_\mathrm{cluster}=305^{+19}_{-25}\,\mathrm{kpc}\) in the fiducial solution. \textsc{ORFEO} thus provides a quantitative dynamical counterpart to the morphology-based picture, extending it to the motion of NGC~4569 through the Virgo ICM. Combining the reconstructed orbit with the adopted orientation of the NGC~4569 disc also yields the wind--disc angle as a derived, rather than independently fitted, quantity; its evolution for the fiducial solution is presented in Appendix~\ref{app:wind_disc_angle}.

\section{Conclusions}
\label{sec:conclusions}

We have presented the first application of \textsc{ORFEO} to NGC~4569, using the resolved spatial and LOS-velocity structure of its H\(\alpha\) tail to reconstruct the recent galaxy trajectory and tail chronology. Combined with the adopted Virgo Cluster~A potential, ICM density profile, and gas surface-density constraints, the reconstruction yields the corresponding ram-pressure history for an assumed galaxy LOS distance. Our main results are as follows.

\begin{enumerate}
\item Placing NGC~4569 at the canonical Virgo Cluster~A distance, \(d_0=16.5\,\mathrm{Mpc}\), \textsc{ORFEO} reconstructs a coherent phase-space sequence for the ionised tail, reproducing both its projected orientation and the monotonic LOS-velocity trend with distance from the galaxy. The fitted tail age is \(T_\mathrm{tail}=250^{+37}_{-19}\,\mathrm{Myr}\), while the ram-pressure maximum occurred at \(t_{P_\mathrm{ram},\mathrm{peak}}=-163^{+11}_{-15}\,\mathrm{Myr}\), close to the reconstructed nearest approach to the Virgo Cluster~A centre at \(305^{+19}_{-25}\,\mathrm{kpc}\). The negative value of \(t_{P_\mathrm{ram},\mathrm{peak}}\) places the fiducial solution after peak ram pressure, while the ordering \(P_\mathrm{ram,peak}>P_\mathrm{ram}(-T_\mathrm{tail})>P_\mathrm{ram}(0)\) shows that the present-day ram pressure
is lower than at the model epoch of the oldest fitted tail tracer.

\item The LOS-distance scan does not determine a unique distance for NGC~4569, but separates comparably fitting reconstructions into two evolutionary families. Low-distance solutions, \(d_0\lesssim16.6\,\mathrm{Mpc}\), place the galaxy after peak ram pressure. At larger distances, the reconstruction moves onto a pre-peak branch, in which the ram-pressure maximum lies in the future and the present-day interaction is still strengthening. The transition occurs around \(d_0\simeq16.8\)--\(17.1\,\mathrm{Mpc}\) and is identified most robustly by the sign change of \(t_{P_\mathrm{ram},\mathrm{peak}}\), although the detailed chronology of the high-distance branch is less stable.

\item An external Gunn--Gott stripping-threshold comparison provides additional physical discrimination between these comparably fitting orbital families and favours the low-distance branch. At the \(1\sigma\) level, \(P_{\mathrm{ram,peak}}\) remains compatible with stripping at the observed disc truncation radius up to \(d_0=16.8\,\mathrm{Mpc}\), while \(P_\mathrm{ram}(-T_\mathrm{tail})\) remains compatible up to \(d_0=16.5\,\mathrm{Mpc}\); \(P_\mathrm{ram}(0)\) remains below the threshold throughout the scan. At higher distances, the compatibility weakens and, beyond the transition, the reconstructed ram-pressure maximum lies in the future, making these solutions less natural as an explanation for a disc truncation independently dated to the past. The low-distance branch therefore provides the preferred post-stripping interpretation of NGC~4569.

\item The favoured low-distance reconstruction places previous constraints on NGC~4569 within a common dynamical picture. Its peak-pressure epoch complements the \(\sim100\,\mathrm{Myr}\) disc-truncation and stellar-population constraint of \citet{Boselli06}, while the \(\sim500\,\mathrm{Myr}\) deceleration timescale estimated by \citet{Sun2026} for the extended H\(\alpha\) tail provides a more direct comparison with \(T_\mathrm{tail}\). The same reconstruction quantitatively recovers the mostly west-to-east projected motion inferred from the H\(\alpha\) morphology of \citet{Boselli16a}, while the post-stripping model of \citet{Vollmer04} provides independent supporting context.

\end{enumerate}

The method is not intrinsically restricted to H\(\alpha\): resolved \HI{} tails provide the same basic observables---projected position and LOS velocity---while directly constraining the gas column density entering the drag model. Other gas phases could also be used if their large-scale kinematics trace the stripped wake and an appropriate effective surface-density prescription can be defined. The minimum useful tail extent is not universal, because the constraining power depends on the spatial and velocity baseline, tracer sampling, and observational uncertainties; controlled truncation-and-recovery tests will be needed to quantify this limit for broader applications of \textsc{ORFEO}.

This first application shows that extended ionised tails can be used as dynamical tracers of the recent interaction histories of cluster galaxies. Despite its simplified treatment of the stripped gas, \textsc{ORFEO} reconstructs the orbit, tail chronology, and ram-pressure evolution of NGC~4569 in only \(\sim10\)--\(11\,\mathrm{min}\) of wall-clock time. These reconstructed histories can provide physically motivated constraints for future tailored simulations and, when applied to larger samples, a statistical view of where, when, and under which ICM conditions gas removal occurs in clusters.

\begin{acknowledgements}
We are grateful to the whole CFHT team who assisted us in the preparation and in the execution of the observations and in the calibration and data reduction.
We acknowledge financial support from ``Programme National de Cosmologie and Galaxies" (PNCG) funded by CNRS/INSU-IN2P3-INP, CEA and CNES, France.
This research has made use of the NASA/IPAC Extragalactic Database (NED) 
which is operated by the Jet Propulsion Laboratory, California Institute of 
Technology, under contract with the National Aeronautics and Space Administration
and of the GOLDMine database (http://goldmine.inaf.it/) \citep{Gavazzi03}.
M. S. acknowledges the support from the NSF grant 2407821.
\end{acknowledgements}

\bibliography{Biblio_cleaned}

\appendix
\section{Gaseous tail modelling}\label{App:tail_validation}
\subsection{Tail construction}\label{app:tail_const}
This section details the numerical implementation of the one-dimensional orbital-spine model introduced in Sect.~\ref{sec:Gaseous_tail}.
Let \(\mathbf{x}_\mathrm{gal}(t)\) denote the galaxy orbit. We define the signed arc-length coordinate along this trajectory as
\begin{equation}
\gamma(t) = \int_0^t \left\| \frac{d\vec{x}_{\text{gal}}}{dt'} \right\| dt'.
\end{equation}
Writing the same trajectory as \(\mathbf{x}(\gamma)\), the position of a gas parcel is \(\mathbf{x}_\mathrm{gas}(t)=\mathbf{x}[\gamma_\mathrm{gas}(t)]\), where \(\gamma_\mathrm{gas}(t)\) evolves under drag and gravity. The total acceleration is decomposed as
\begin{equation}
\vec{a}_{\text{gas}} = \vec{a}_{\text{gal}} - \frac{\vec{v}_{\text{gas}}}{\tau} + \vec{a}_{\text{track}},
\end{equation}
where \( \vec{a}_{\text{gal}} \) traces the gravitational field of the cluster, the second term accounts for ram-pressure drag, and \( \vec{a}_{\text{track}} \) is the acceleration required to keep the gas confined to the orbit. Since we impose geometric confinement, we do not compute \( \vec{a}_{\text{track}} \) explicitly; instead, we retain only the tangential component of the acceleration.

Defining the unit tangent vector along the orbit as
\(\hat{t}(t) = \frac{d\vec{x}_{\text{gal}}/dt}{\left\| d\vec{x}_{\text{gal}}/dt \right\|},\)
the projected gravitational acceleration induced by the cluster gravitational potential becomes
\(a_{\parallel}(t) = \vec{a}_{\text{gal}}(t) \cdot \hat{t}(t),
\) which we interpolate as a function of arc-length. The motion of the gas is then governed by the one-dimensional differential equation
\begin{equation}\label{eq:diff_gamma}
\frac{d^2 \gamma_{\text{gas}}}{dt^2} = a_{\parallel}(\gamma_{\text{gas}}) - \frac{1}{\tau} \frac{d\gamma_{\text{gas}}}{dt}.
\end{equation}
Here \(\tau(t;t_\mathrm{strip})\) is the ram-pressure drag
timescale defined in Eq.~\eqref{eq:tau}. For each parcel,
the assigned surface density remains fixed, while the local
ICM density and parcel speed vary along its trajectory,
making \(\tau\) time-dependent. We integrate
Eq.~\eqref{eq:diff_gamma} using a leapfrog scheme.
At each time step, the gas velocity is reconstructed as
\begin{equation}
\vec{v}_{\text{gas}}(t) = \frac{d\gamma_{\text{gas}}}{dt} \cdot \hat{t}(\gamma_{\text{gas}}),
\end{equation}
ensuring that the motion remains tangent to the orbit. This reduces the dynamics of each stripped parcel to a single degree of freedom.

An ensemble of parcels is evolved from its assigned \(t_\mathrm{strip}\) to \(t=0\) using the initial conditions of Eq.~\eqref{eq:gas_initial_conditions}. Present-day positions and velocities are obtained by interpolating the orbit at \(\gamma_\mathrm{gas}(0)\), defining the model tail used for comparison with simulations or observations.

\subsection{Benchmark of the effective drag prescription}
\label{app:Benchmark_drag}
\begin{figure}
    \centering
    \includegraphics[width=\linewidth]{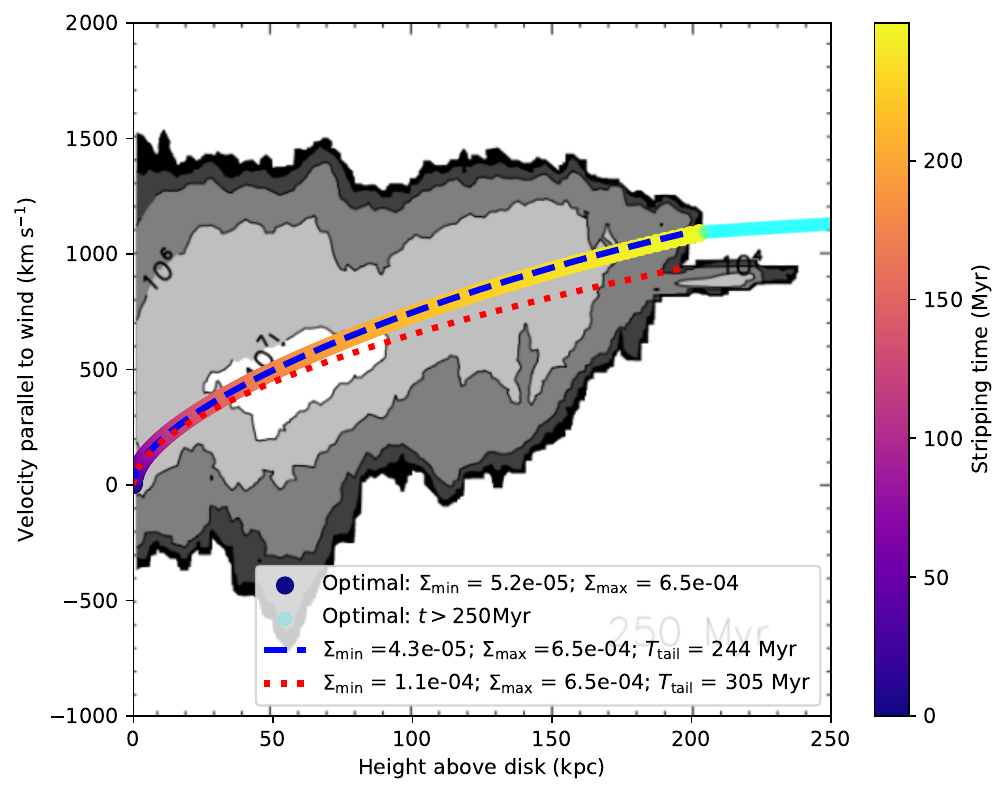}
    \caption{Comparison between the simulation of \citetalias{Tonnesen10} (density-weighted gas distribution, background contours) and reconstructed gas tails at $t=250\,\mathrm{Myr}$. The solid curve shows the ``optimal'' reconstruction, with colours encoding the lookback stripping time of each gas parcel; the cyan extension marks the extrapolation beyond the simulation epoch. The dashed and dotted curves correspond to the lower and upper bounds on the assumed minimum tail surface density, $\Sigma_{\mathrm{tail,min}}$, extrapolated to the time best matching the length of the tail $T_\mathrm{tail}$, indicated in the legend.}
    \label{fig:Drag_timescale}
\end{figure}

We benchmark the effective post-release drag prescription against the \(T_\mathrm{min}=300\,\mathrm{K}\) hydrodynamical simulation of \citet{Tonnesen10}, hereafter \citetalias{Tonnesen10}. The simulation adopts a constant ICM wind with \(v_\mathrm{ICM}=1413\,\mathrm{km\,s^{-1}}\) and \(P_\mathrm{ram}=6.4\times10^{-12}\,\mathrm{dyn\,cm^{-2}}\). We reproduce this configuration with a straight-line trajectory at \(v=-v_\mathrm{ICM}\) and compare the resulting longitudinal tail structure with the simulation. As in Sect.~\ref{sec:Gaseous_tail}, we assign parcel surface densities between a dense-gas anchor, \(\Sigma(R_\mathrm{strip})\), and a minimum tail value, \(\Sigma_\mathrm{tail,min}\). Under the constant-wind conditions of \citetalias{Tonnesen10}, the ICM density is fixed, while the drag timescale in Eq.~\eqref{eq:tau} varies through the assigned surface density and the instantaneous parcel velocity. We fix \(\Sigma(R_\mathrm{strip})=6.5\times10^{-4}\,\mathrm{g\,cm^{-2}}\), corresponding to \(N_\mathrm{H}\simeq3\times10^{20}\,\mathrm{cm^{-2}}\) and consistent with \citetalias{Tonnesen10}, and compare three values of \(\Sigma_\mathrm{tail,min}\). The ``optimal'' case, \(5.2\times10^{-5}\,\mathrm{g\,cm^{-2}}\), is calibrated to match both the simulated tail extent and its \(250\,\mathrm{Myr}\) age, while \(4.3\times10^{-5}\) and \(1.1\times10^{-4}\,\mathrm{g\,cm^{-2}}\) bracket the observationally motivated range adopted for NGC~4569 and test the response to stronger and weaker effective drag, respectively.
Figure~\ref{fig:Drag_timescale} compares the \citepalias{Tonnesen10} simulation at \(t=250\,\mathrm{Myr}\) with the reconstructed tails. All three prescriptions reproduce the velocity gradient of the dense inner tail, with increasing differences towards larger distances while remaining within the characteristic velocity scatter of the simulated wake. The calibrated case matches the \(250\,\mathrm{Myr}\) tail age by construction; its independent test is therefore the agreement of the simplified drag model with the simulated velocity--distance structure. The lower- and upper-density bounds yield \(T_\mathrm{tail}=244\) and \(305\,\mathrm{Myr}\), respectively, corresponding to a spread of \(\simeq60\,\mathrm{Myr}\). Appendix~\ref{app:Likelihood} tests separately how uncertainty in the adopted tail-density scale propagates into the inverse recovery.

\subsection{Scope of the gaseous-tail model}
\label{app:tail_scope}

\textsc{ORFEO} is an inverse model of the large-scale phase-space spine of the observed tail, rather than a forward model of gas removal from the disc. Parcels are sampled at assigned stripping times \(t_\mathrm{strip}\leq0\), independently of a ram-pressure threshold or stripping-efficiency criterion. Thus, \(t_\mathrm{strip}\) labels a model parcel along the present-day tail rather than predicting the physical instant at which that gas became unbound. Accordingly, \(T_{\rm tail}\) is the age of the oldest fitted H\(\alpha\) parcel and need not coincide with the onset of gas removal.

At \(t_\mathrm{strip}\), each parcel is initialised at the galaxy position with the galaxy bulk orbital velocity. The model does not include finite launch radius or azimuth, an evolving stripping radius, motion through the galaxy potential during detachment, or disc rotation. Parcel release is also independent of wind--disc inclination and does not distinguish between direct ram-pressure displacement, edge stripping, or Kelvin--Helmholtz removal \citep{RoedigerHensler2005}. The wind--disc angle derived in Appendix~\ref{app:wind_disc_angle} therefore characterises the reconstructed geometry but does not enter the release prescription.

After release, parcels remain confined to the one-dimensional orbital spine and evolve under projected cluster gravity and effective ram-pressure drag.
The galaxy's own gravitational attraction is neglected in this post-release evolution. Where its component along the spine opposes drag, it reduces the net gas deceleration.
This effect may be partly represented by a larger effective surface density and hence a longer drag timescale, although the prescribed coupling cannot generally reproduce the full time dependence of the omitted force. Tail width, lateral accelerations, turbulence, filamentation, expansion or fragmentation, multiphase evolution, and ICM bulk motions are not resolved. Unresolved changes in scalar post-release gas--ICM coupling are represented through the effective \(C_\mathrm{d}/\Sigma_\mathrm{gas}\) prescription, whereas coherent launch-side offsets or systematic transverse displacements are not. Such systematic displacements of the collective
tail spine would affect the recovered orbit and chronology,
which are therefore conditional on the adopted orbital-spine
approximation. Quantifying these effects requires calibrating
departures from the galaxy trajectory against hydrodynamical
simulations of curved galaxy orbits.

Observed H\(\alpha\) regions are associated independently with the nearest location on the model tail in \((\mathrm{RA},\mathrm{DEC},cz_\mathrm{obs})\) through Eq.~\eqref{eq:model_data_association}. No requirement is imposed that every model segment be detectable in H\(\alpha\), and no hard association distance is applied; mismatches are instead penalised continuously by their spatial and kinematic residuals. Gas older than the oldest fitted H\(\alpha\) tracer may therefore be absent from the observed sequence. The model also does not separate hydrodynamical from possible tidal contributions, and its orbital interpretation applies to the large-scale component associated with the galaxy--ICM interaction.

\section{Likelihood validation}\label{app:Likelihood}
The validation suite presented in this appendix was developed using the earlier 83-region version of the observational catalogue. Accordingly, the controlled mock is sampled with 83 tracers, and the \(t_{\rm scale}\) diagnostic in Appendix~\ref{App:tscale_optimisation} illustrates the corresponding 83-region reconstruction. The results reported in the main text have instead been recomputed using the final 94-region catalogue. The reconstruction algorithm and validation procedure are unchanged.
\begin{figure}
  \centering
  \includegraphics[width=\columnwidth]{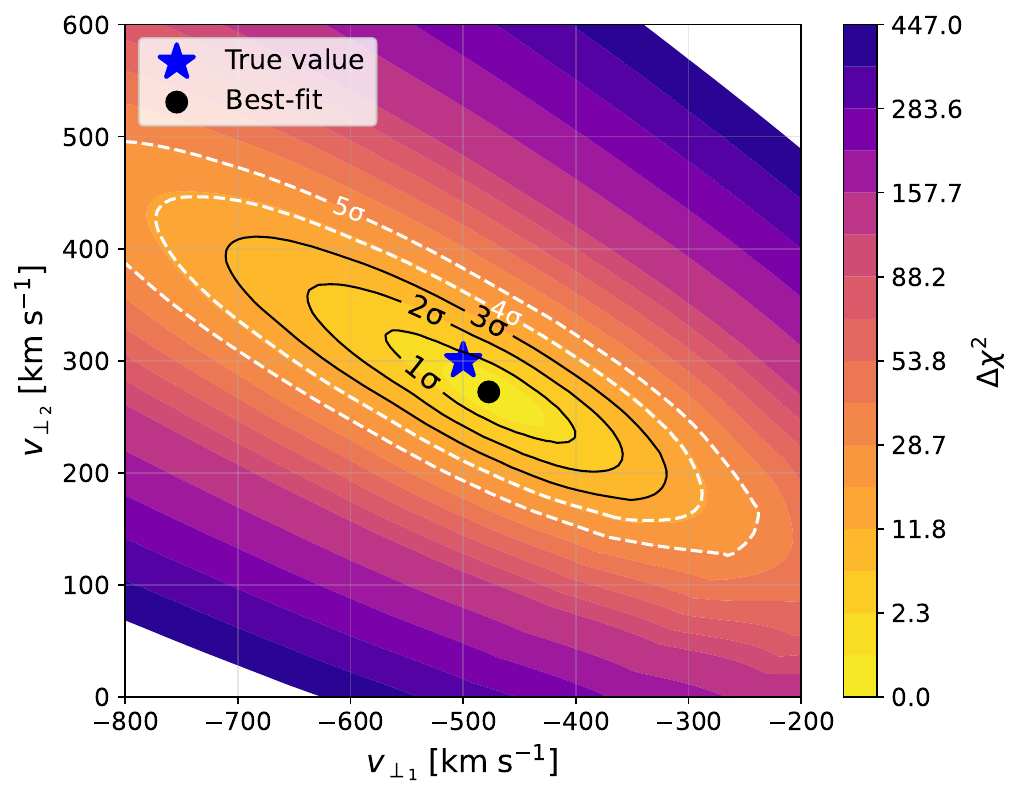}
  \caption{\(\Delta\chi^2\) surface of the transverse velocity components $(v_{\perp,1},v_{\perp,2})$. The best-fit solution is marked by a black dot; the true input by a blue star. Black solid lines indicate the nominal \(1\)--\(3\sigma\) two-parameter contours, while white dashed lines extend to \(5\sigma\). The colour scale shows $\Delta\chi^2$ relative to the best fit.}
  \label{fig:velocity_contours}
\end{figure}
\begin{table}[ht]
\centering
\begin{tabular}{lcc}
\hline
{Quantity} & {$\Delta$} & {$\sigma$} \\
\hline
$v_{\perp,1}$ (km s$^{-1}$) & -23 & 58 \\
$v_{\perp,2}$ (km s$^{-1}$)& 28 & 35 \\
$T_\mathrm{tail}$ (Myr)& 15 & 17  \\
$T_{P_{\rm ram,\max}} (Myr)$ & 1 & 7\\
\hline
\end{tabular}
\caption{Recovery of the injected quantities in the controlled mock. We define \(\Delta\equiv q_{\rm true}-q_{\rm best}\); \(\sigma\) denotes the corresponding nominal \(1\sigma\) uncertainty.}\label{tab:Toy_results}
\end{table}
We test the inverse-reconstruction framework on a controlled mock with known orbital parameters and tail chronology. The purpose is not to reproduce the full complexity of the observations, but to determine whether the fitting procedure recovers the injected quantities from a sparsely sampled tail with realistic spatial and kinematic scatter.

We describe the Virgo mass profile with a Navarro--Frenk--White (NFW) profile with \(R_{200}=1623.3\,\mathrm{kpc}\), \(M_{200}=5.3\times10^{14}\,M_\odot\), and \(R_\mathrm{s}=112\,\mathrm{kpc}\). For the ICM, we adopt the single-\(\beta\) electron-density profile of \citet{Simionescu2017},
\[
n_{\mathrm{e,ICM}}(r) =
n_{\mathrm e,0}
\left[
1+\left(\frac{r}{r_\mathrm c}\right)^2
\right]^{-3\beta/2},
\]
with \(n_{\mathrm e,0}=3.13\times10^{-2}\,\mathrm{cm^{-3}}\), \(r_\mathrm c=13\,\mathrm{kpc}\), and \(\beta=0.47\). The gas surface-density prescription is set by the surface density at the stripping radius and by an observationally motivated minimum tail surface density. Specifically, we adopt \(N_{\rm H,gas}(R_{\rm strip})=1\times10^{21}\,\mathrm{cm^{-2}}\) as the dense-gas anchor at the present-day disc truncation radius and \(N_{\rm H,tail,min}=3.2\times10^{19}\,\mathrm{cm^{-2}}\) as the adopted minimum effective column density
in the mock tail. These values are consistent with the dense-disc reference and the adopted tail reference range discussed in Sect.~\ref{Sec:data_NGC} and are converted to surface density through \(\Sigma_{\rm gas}=\mu m_{\rm p}N_{\rm H}\), with \(\mu=1.3\).
The resulting tail is truncated at a maximum lookback time of \(300\,\mathrm{Myr}\) to emulate a system that began stripping at a fixed epoch in the past. It is then converted into observable quantities---\(RA\), \(DEC\), projected distance from the galaxy nucleus (\(D_\mathrm{nucleus}\)), and LOS velocity (\(cz_\mathrm{obs}\))---and perturbed to reproduce the effective spatial and kinematic dispersion observed around the mean tail trends. The different dispersions adopted in \(RA\) and \(DEC\) reflect the anisotropic width of the observed tail in these projections. Gaussian noise is added with standard deviations of
\(0.0001\,\mathrm{h}\) in RA, \(0.1^\circ\) in Dec,
and \(120\,\mathrm{km\,s^{-1}}\) in velocity. A random subset of 83 tail points is then selected, matching the sampling density of the earlier observational catalogue on which this validation was developed.

To test our model, we apply our likelihood framework to this perturbed dataset, scanning the $(v_{\perp,1},v_{\perp,2})$ parameter space while fixing the galaxy position. The goal is threefold: assess whether the tail morphology and kinematics are reproduced; test how well the time-dependent ram pressure is recovered; and determine whether the input initial conditions and derived stripping timescales are retrieved within realistic uncertainties. Figure~\ref{fig:velocity_contours} shows the \(\Delta\chi^2\) surface in the $(v_{\perp,1}, v_{\perp,2})$ plane. The black dot marks the best-fitting value, while the nominal \(1\)--\(3\sigma\) two-parameter contours are shown as black solid lines; the \(4\)--\(5\sigma\) contours are shown as white dashed lines. For this realisation, the recovered velocities are $v_{\perp,1}=-477\pm58~\mathrm{km\,s^{-1}}$ and $v_{\perp,2}=272\pm35~\mathrm{km\,s^{-1}}$, consistent with the true values (-500, 300) $\mathrm{km\,s^{-1}}$ used to generate the tail, marked by a blue star.

\begin{figure}
  \centering
  \includegraphics[width=\columnwidth]{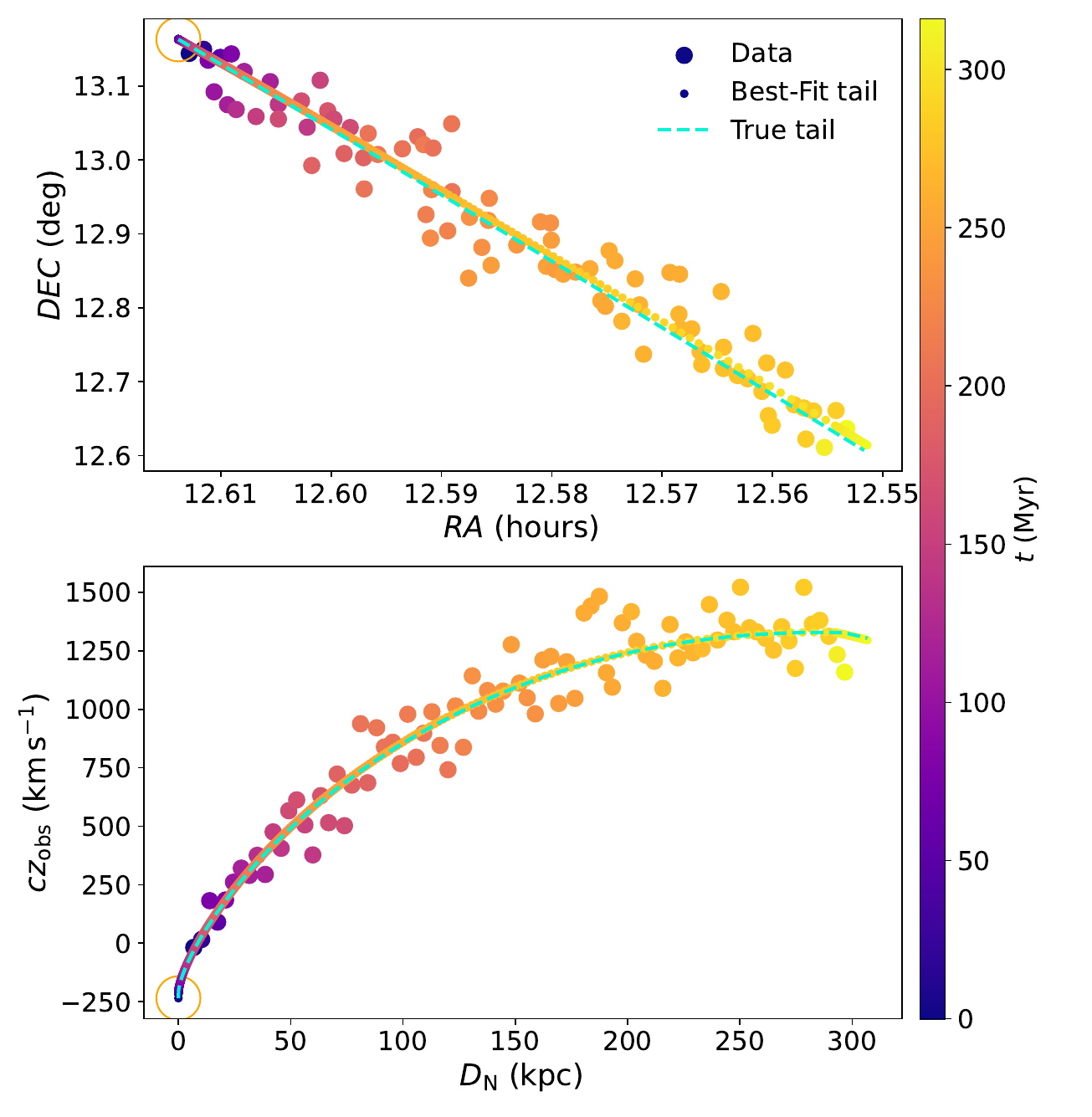}
  \caption{Validation of the reconstructed trajectory against the toy model. \emph{Top panel}: distribution of mock tail points in RA–DEC. \emph{Bottom panel}: LOS velocity as a function of projected distance from the galaxy nucleus. The dashed light-blue line shows the unperturbed model used to generate the data. The solid line shows the best-fitting reconstructed model tail, colour-coded by lookback time.}
  \label{fig:orbit_validation}
\end{figure}

\begin{figure}
  \centering
  \includegraphics[width=\columnwidth]{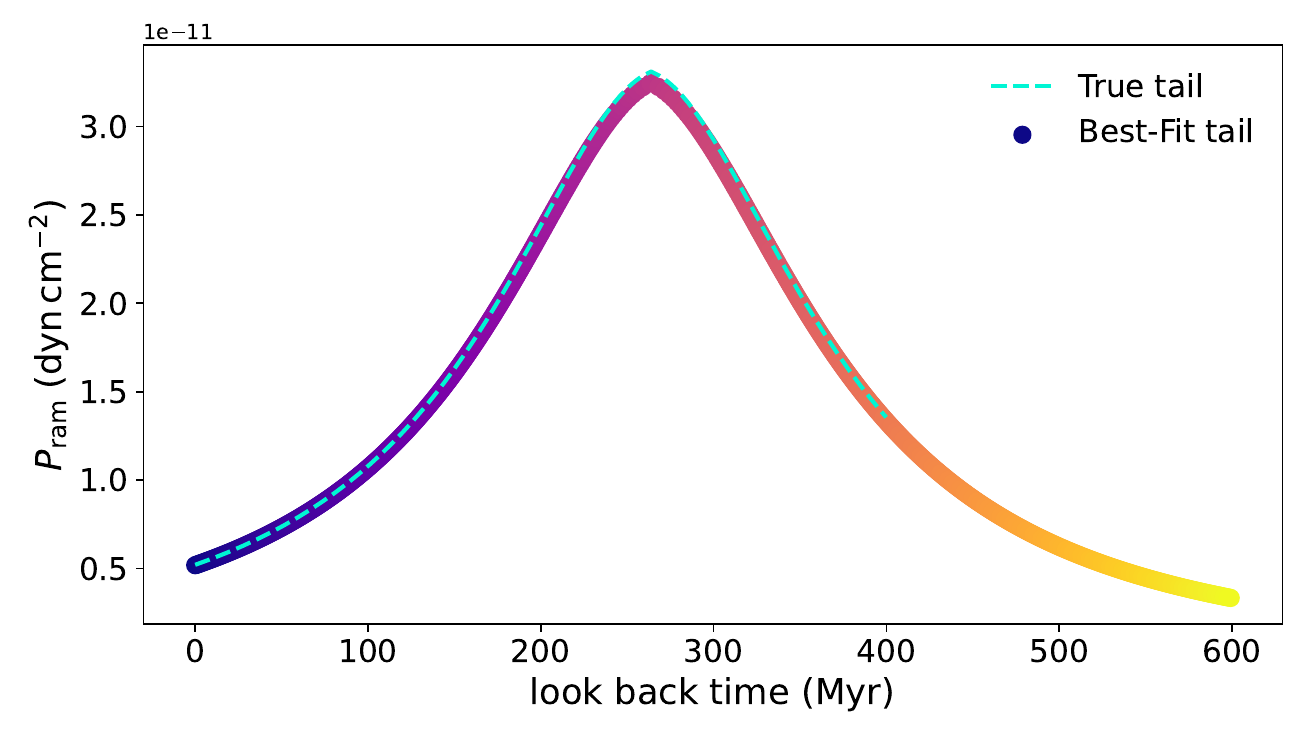}
  \caption{Evolution of ram pressure as a function of lookback time for the reconstructed (solid line) and input (dashed line) orbits. Ram pressure is computed from the galaxy’s velocity in the cluster frame and the local ICM density along the orbit.}
  \label{fig:pram_evolution}
\end{figure}

Completing the picture, Fig.~\ref{fig:orbit_validation} shows the best-fit model for the tail compared with the mock dataset. The data points (coloured dots) are shown in the $RA--DEC$ plane (top) and as LOS velocity versus projected distance from the nucleus (bottom). Colours encode the model lookback time assigned to each mock tracer by the reconstruction. The best-fit reconstructed curve (solid line, colour-coded by lookback time) closely follows the unperturbed generator model (light-blue dashed line), reproducing both the on-sky morphology and the $v_{\rm los}$–$D_\mathrm{nucleus}$ relation. The maximum reconstructed lookback time recovers the injected tail age of \(300\,\mathrm{Myr}\) to within \(15\,\mathrm{Myr}\). Going beyond the pure orbital information, Fig.~\ref{fig:pram_evolution} shows the retrieved ram-pressure history, computed as $P_{\rm ram}(t)=\rho_{\rm ICM}[r(t)]\,v^2(t)$, for the injected (dashed) and reconstructed (solid) orbits. To test the orbital extrapolation beyond the interval directly sampled by the mock tail, we extend the reconstructed orbit beyond \(300\,\mathrm{Myr}\). The reconstructed curve follows the injected one closely over the full interval, with only a mild underestimation near the peak. This agreement supports the reliability of the reconstructed ram-pressure history under the controlled assumptions of the test.

In addition to the initial condition, we focus on two derived timescales: the tail age, $T_\mathrm{tail}$, defined as the maximum lookback time assigned to any tail point, and the time of peak ram pressure, $T_{P_{\rm ram,\max}}$. These are not fit parameters, but are computed \emph{a posteriori} from the best-fit trajectory. Table~\ref{tab:Toy_results} summarises the deviations from the true values and their uncertainties. 
For this realisation, the recovered \(T_\mathrm{tail}\) and \(T_{P_\mathrm{ram,max}}\) differ from their injected values by \(15\) and \(1\,\mathrm{Myr}\), respectively, both within the nominal uncertainties reported in Table~\ref{tab:Toy_results}. Together with the close match in Fig.~\ref{fig:pram_evolution}, this shows that the reconstruction recovers both the injected timescales and the ordering of the tail tracers despite their independent nearest-point association.

\subsection{Sensitivity to input gas density}

Whereas Appendix~\ref{app:Benchmark_drag} benchmarks the effective drag prescription against an independent hydrodynamical simulation, here we test how an incorrect \(\Sigma_\mathrm{tail,min}\) propagates into the inverse recovery of the orbit and chronology. We consider \(N^\mathrm{low}_\mathrm{H,tail,min}=2\times10^{19}\,\mathrm{cm^{-2}}\) and \(N^\mathrm{high}_\mathrm{H,tail,min}=5\times10^{19}\,\mathrm{cm^{-2}}\), bracketing the value used to generate the mock. As in the validation against the \citetalias{Tonnesen10} simulation (Appendix~\ref{App:tail_validation}), and for the same column densities, we find that underestimating the density (blue dashed line) does not degrade the fit, which still reproduces both the morphology in the $RA--Dec$ plane and the velocity gradient. In contrast, overestimating the density (red dotted line) leads to a visibly poorer match, particularly at large distances from the nucleus—consistent with the trend shown in Fig.~\ref{fig:Drag_timescale}.
\begin{figure}
  \centering
  \includegraphics[width=\columnwidth]{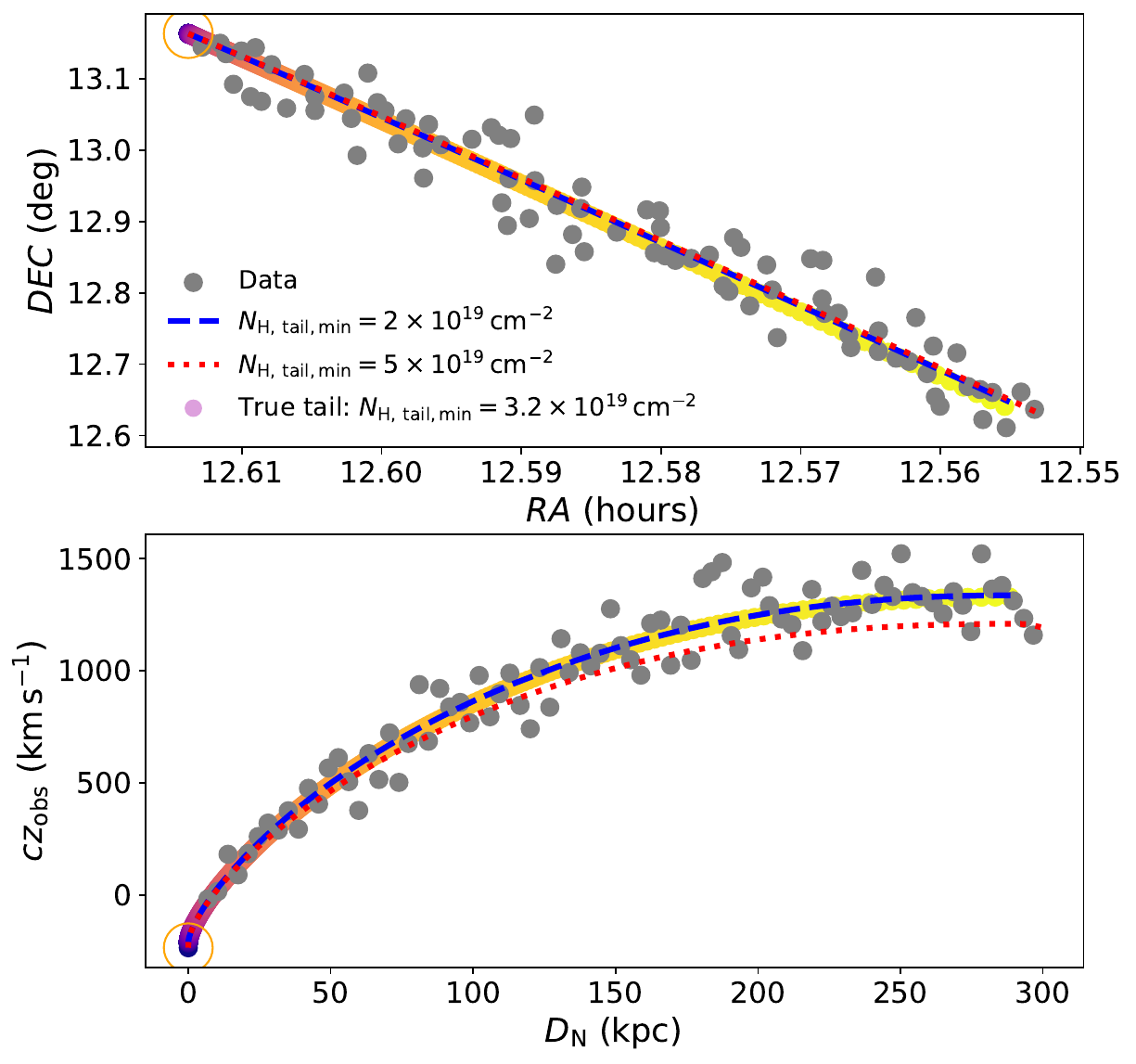}
  \caption{Reconstructed mock tail for different assumed minimum tail column densities. \emph{Top panel}: distribution of mock tail points in the \(RA--DEC\) plane. \emph{Bottom panel}: LOS velocity as a function of projected distance from the galaxy nucleus. Grey points show the mock dataset. The blue dashed and red dotted curves show the best-fitting model tails obtained assuming \(N_{\rm H,tail,min}=2\times10^{19}\,\mathrm{cm^{-2}}\) and \(5\times10^{19}\,\mathrm{cm^{-2}}\), respectively. Coloured points show the input model for reference.}
  \label{fig:orbit_validation_sigma}
\end{figure}

We quantify the relative agreement of the \textsc{ORFEO} predictions with the input dataset using the same reduced \(\chi^2\)-like statistic adopted in the main analysis. Using \(\nu=2N-2=164\) as a common normalisation for the 83-point mock, we obtain
\(
\chi^2_\mathrm{true}/\nu = 0.35 \quad (\text{reference model}), \quad
\chi^2_\mathrm{low}/\nu = 0.35 \quad (\text{low density}), \quad
\chi^2_\mathrm{high}/\nu = 0.57 \quad (\text{high density}).
\)
These values are used as comparative indicators of fit quality rather than as an exact \(\chi^2\) probability calibration.
The reference and low-density cases are indistinguishable according to this diagnostic, whereas the high-density assumption produces a substantially poorer fit. In this controlled mock, the diagnostic distinguishes the reference and low-density prescriptions from the incompatible high-density case.

To further assess whether an incorrect assumption on $\Sigma_\mathrm{tail,min}$ biases the relevant quantities estimated by \textsc{ORFEO}, we perform the complete likelihood analysis for the three density cases, focusing on $(v_{\perp,1},v_{\perp,2})$, $T_\mathrm{tail}$, and $T_{P_{\rm ram,\max}}$. The results are summarised in Table~\ref{tab:Toy_results_sigma}, expressed as deviations from the expected values $\Delta$. For comparison, the corresponding results obtained with the true density are reported in Table~\ref{tab:Toy_results}.
\begin{table}[]
\centering
\begin{tabular}{lcccc}
\hline
{Quantity} & {$\Delta_\mathrm{low}$} & {$\sigma_\mathrm{low}$} & {$\Delta_\mathrm{high}$} & {$\sigma_\mathrm{high}$}\\
\hline
$v_{\perp,1}$ (km s$^{-1}$) & -8 & 63 & 105 & 40 \\
$v_{\perp,2}$ (km s$^{-1}$) & -5 & 37 & -81  & 24  \\
$T_\mathrm{tail}$ (Myr)& -13  & 17  & 108 & 17 \\
$T_{P_{\rm ram,\max}}$ (Myr) &  0 &  7 & -12 & 15 \\
\hline
\end{tabular}
\caption{Likelihood validation for the two bounding tail-density assumptions. The low- and high-density cases correspond to \(N_{\rm H,tail,min}=2\times10^{19}\,\mathrm{cm^{-2}}\) and \(5\times10^{19}\,\mathrm{cm^{-2}}\), respectively.} Residuals (\(\Delta\)) are computed with respect to the true input values, whose recovery is reported in Table~\ref{tab:Toy_results}.\label{tab:Toy_results_sigma}
\end{table}

The lower-density assumption leaves the recovered orbit and chronology consistent with the injected values within the nominal \(1\sigma\) uncertainties. The higher-density case instead produces significant deviations, most notably \(\lvert\Delta T_{\rm tail}\rvert=108\,\mathrm{Myr}\). Together with its poorer fit statistic, this shows that the high-density prescription does not recover the known orbit and chronology of the controlled mock. This does not imply that a comparable separation between the age of the oldest observable tail material and peak ram pressure would be intrinsically unphysical in a real system, since older stripped gas may no longer be detectable. Here, however, the imposed onset and orbital history are known, so the chronology provides an additional diagnostic of model mismatch. In the analysis of NGC~4569, we therefore explore different column-density assumptions and assess both their relative fit quality and the physical plausibility of the reconstructed chronology.

\subsection{Sensitivity to the stripping timescale}
\label{App:tscale_optimisation}
\begin{figure}
  \centering
  \includegraphics[width=\columnwidth]{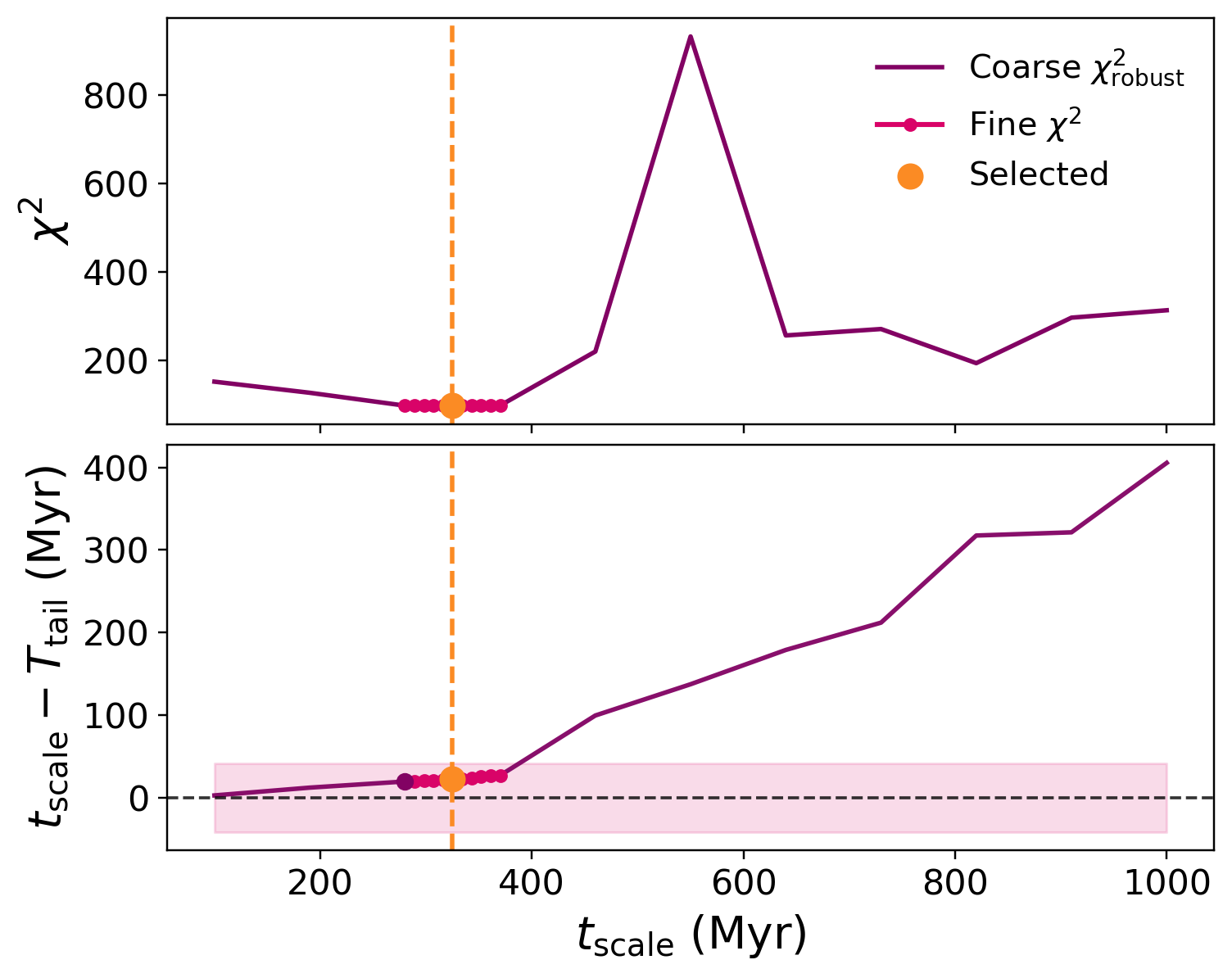}
     \caption{Diagnostic plots for the \(t_{\rm scale}\) optimisation, illustrated using the earlier 83-region reconstruction of NGC~4569. \emph{Top panel}: coarse (purple) and fine (magenta) exploration of \(\chi^2_{\rm robust}\) as a function of the trial \(t_{\rm scale}\). The vertical dashed line marks the adopted solution. \emph{Bottom panel}: difference between the trial effective surface-density timescale and the reconstructed tail age. The magenta band shows the combined within-model uncertainty at each scan point, and the horizontal dashed line marks exact consistency, \(t_{\rm scale}=T_{\rm tail}\).}\label{fig:tscale_diagnostic}
\end{figure}

The effective gas-surface-density prescription adopted by \textsc{ORFEO} requires a trial effective surface-density evolution timescale \(t_{\rm scale}\), which sets the temporal normalisation of Eq.~\eqref{eq:sigma_evol}. Since \(t_\mathrm{scale}\) is not known \emph{a priori}, it is selected through an iterative scan coupled to the orbit reconstruction. For each trial \(t_\mathrm{scale}\), \textsc{ORFEO} returns the best-fitting orbital parameters, fit statistic, and reconstructed \(T_\mathrm{tail}\). Because \(t_\mathrm{scale}\) sets the interval over which the assigned surface-density sequence reaches the tail-density scale, we retain solutions for which \(t_\mathrm{scale}\) and \(T_\mathrm{tail}\) are consistent within their uncertainties. The requirement \(t_{\rm scale}\simeq T_{\rm tail}\) is therefore an internal consistency criterion of the adopted surface-density parametrisation, rather than an independent physical measurement of the stripping onset.
The optimisation is performed in two stages. We first carry out a coarse scan over a broad range of plausible \(t_{\rm scale}\) values. At each value of \(t_{\rm scale}\), we minimise \(\chi^2\) over the two transverse velocity components and assess the numerical quality of the resulting minimum on a local \(8\times8\) grid spanning \(\pm3\) Hessian standard deviations. To prevent poorly resolved or boundary-truncated minima from driving the selection, we define
\begin{equation}
\chi^2_{\rm robust}(t_{\rm scale})=\chi^2(t_{\rm scale})+P(t_{\rm scale}),
\end{equation}
with
\begin{equation}
P =
20I_{\rm H}+40I_{\rm G}+20I_{\rm edge}
+30\max\left[0,\frac{2.30-B}{2.30}\right]
+10\max\left[0,\frac{5-N}{5}\right].
\end{equation}
Here \(I_{\rm H}\), \(I_{\rm G}\), and \(I_{\rm edge}\) flag an invalid Hessian, failure of the local-grid calculation, and intersection of the \(\Delta\chi^2\leq2.30\) region with the grid boundary, respectively; \(B\) is the minimum boundary value of \(\Delta\chi^2\), and \(N\) is the number of grid points within \(\Delta\chi^2\leq2.30\). The penalty acts only as a numerical-quality safeguard: for accepted, well-resolved solutions \(P=0\), so that \(\chi^2_{\rm robust}=\chi^2\) and their ranking is unchanged. The penalty is not included in the reported goodness-of-fit or confidence intervals. If several local minima are present, we retain the basin with the lowest \(\chi^2_{\rm robust}\) among the numerically accepted solutions for which \(t_{\rm scale}\) remains compatible with \(T_{\rm tail}\) within its local uncertainty estimate. A second, refined scan is then performed within this basin to determine the preferred stripping timescale more precisely.
The uncertainty associated with the input choice of \(t_{\rm scale}\) is propagated into the final reconstructed quantities as an additional systematic contribution, \(\sigma_{\rm sys}\), estimated from the spread of the best-fitting parameters over the \(1\sigma\) interval of the allowed \(t_{\rm scale}\) range. The reported within-model uncertainty is then obtained by combining these terms in quadrature,
\begin{equation}
\sigma_{\rm tot}^2 = \sigma_{\rm fit}^2 + \sigma_{\rm sys}^2 .
\end{equation}
Here \(\sigma_{\rm tot}\) denotes the combined uncertainty within the adopted cluster, ICM, and tail model; uncertainties associated with changing these structural assumptions are not included.

Figure~\ref{fig:tscale_diagnostic} illustrates the \(t_{\rm scale}\) optimisation using the earlier 83-region reconstruction of NGC~4569 at \(d_0=16.5\,\mathrm{Mpc}\) and \(\Sigma^\mathrm{fid}_{\rm tail,min}=7.6\times10^{-5}\,\mathrm{g\,cm^{-2}}\). The figure is included to document the selection procedure; the results reported in Sect.~\ref{sec:fiducial_fixed_distance} are obtained from the final 94-region catalogue. The coarse scan identifies the preferred numerically accepted basin, and the subsequent fine scan refines \(t_{\rm scale}\) within that region. The residual \(t_{\rm scale}-T_{\rm tail}\) confirms that the adopted value satisfies the internal consistency criterion. The uncertainty associated with \(t_{\rm scale}\) is then propagated into the reported ORFEO constraints.

\section{Full \textsc{MG-MAMPOSSt} analysis}\label{app:manposst}

\begin{table*}[]
    \centering
    \small
    \setlength{\tabcolsep}{3pt}
    \caption{Constraints on the cluster parameters from the \textsc{MG-MAMPOSSt} fit for different mass, anisotropy, and galaxy number-density models. The last three rows refer to the full galaxy sample (ALL), while the remaining rows refer to the early-type sample (ET). The last column reports the BIC difference relative to the best-fitting model for each sample (pHer+Her+gOM for ET and Bur+pHer+gOM for ALL). The bold row marks the selected model.}
    \begin{tabular}{c c c c | c c c c c c c}
    \toprule
    $M(r)$ model & $\beta(r)$ model & $N(r)$ model & sample & $r_{200}$ $[\mathrm{Mpc}]$ & $M_{200}$ $[10^{14}\,\mathrm{M}_\odot]$ & $r_\mathrm{s}$ $[\mathrm{Mpc}]$ & $r_\nu$ $[\mathrm{Mpc}]$ & $\mathcal{A}_\infty$ & $\mathcal{A}_0$ & $\Delta\mathrm{BIC}$\\
    \midrule
    Her & gOM & pHer & ET & $1.66^{+0.28}_{-0.29}$ & $5.2^{+2.9}_{-2.4}$ & $0.86^{+0.48}_{-0.75}$ & $0.50^{+0.06}_{-0.12}$ & $1.68^{+1.92}_{-1.24}$ & $1.68^{+1.91}_{-0.92}$ & 0\\[2mm]
    Her & gT & pHer & ET & $1.57^{+0.37}_{-0.21}$ & $5.1^{+3.0}_{-2.4}$ & $0.57^{+0.67}_{-0.49}$ & $0.51^{+0.05}_{-0.13}$ & $1.68^{+1.93}_{-1.24}$ & $1.26^{+1.98}_{-0.90}$ & 0.67\\[2mm]
    Her & gOM & pNFW & ET & $1.51^{+0.27}_{-0.15}$ & $4.5^{+2.0}_{-1.7}$ & $0.68^{+0.39}_{-0.70}$ & $0.18^{+0.04}_{-0.07}$ & $1.69^{+1.92}_{-1.24}$ & $1.29^{+1.94}_{-0.93}$ & 10.52\\[2mm]
    \textbf{NFW} & \textbf{gOM} & \textbf{pHer} & \textbf{ET} & $\boldsymbol{1.63^{+0.32}_{-0.21}}$ & $\boldsymbol{5.4^{+2.9}_{-2.3}}$ & $\boldsymbol{0.36^{+0.32}_{-0.07}}$ & $\boldsymbol{0.50^{+0.12}_{-0.06}}$ & $\boldsymbol{1.70^{+1.86}_{-1.23}}$ & $\boldsymbol{1.34^{+1.96}_{-0.98}}$ & \textbf{0.25}\\[2mm]
    NFW & gT & pHer & ET & $1.57^{+0.40}_{-0.16}$ & $5.4^{+3.1}_{-2.4}$ & $0.23^{+0.43}_{-0.20}$ & $0.50^{+0.12}_{-0.05}$ & $1.72^{+1.90}_{-1.27}$ & $1.34^{+1.87}_{-0.86}$ & 1.14\\[2mm]
    NFW & gOM & pNFW & ET & $1.45^{+0.33}_{-0.24}$ & $4.5^{+1.9}_{-1.6}$ & $0.27^{+0.22}_{-0.24}$ & $0.19^{+0.04}_{-0.08}$ & $1.70^{+1.87}_{-1.24}$ & $1.30^{+1.95}_{-0.94}$ & 10.90\\[2mm]
    Bur & gOM & pHer & ET & $1.71^{+0.23}_{-0.30}$ & $5.3^{+2.8}_{-2.2}$ & $0.36^{+0.07}_{-0.24}$ & $0.50^{+0.12}_{-0.06}$ & $1.62^{+1.91}_{-1.19}$ & $1.36^{+1.91}_{-1.00}$ & 1.00\\[2mm]
    Bur & gT & pHer & ET & $1.58^{+0.34}_{-0.19}$ & $5.2^{+2.8}_{-2.2}$ & $0.17^{+0.15}_{-0.14}$ & $0.50^{+0.12}_{-0.06}$ & $1.66^{+1.91}_{-1.22}$ & $1.30^{+1.95}_{-0.94}$ & 0.94\\[2mm]
    Bur & gOM & pNFW & ET & $1.57^{+0.20}_{-0.19}$ & $4.5^{+1.8}_{-1.5}$ & $0.17^{+0.06}_{-0.06}$ & $0.14^{+0.13}_{-0.10}$ & $1.61^{+1.87}_{-1.16}$ & $1.41^{+1.98}_{-1.06}$ & 11.70\\[2mm]
    \midrule
    Her & gOM & pHer & ALL & $1.85^{+0.26}_{-0.24}$ & $7.4^{+3.2}_{-2.7}$ & $1.04^{+0.75}_{-0.64}$ & $0.57^{+0.07}_{-0.09}$ & $1.51^{+1.79}_{-1.02}$ & $1.24^{+1.72}_{-0.87}$ & 0.43\\[2mm]
    NFW & gOM & pHer & ALL & $1.86^{+0.30}_{-0.26}$ & $7.5^{+3.8}_{-3.0}$ & $0.54^{+0.61}_{-0.45}$ & $0.57^{+0.07}_{-0.09}$ & $1.53^{+1.80}_{-1.02}$ & $1.22^{+1.74}_{-0.87}$ & 0.51\\[2mm]
    Bur & gOM & pHer & ALL & $1.85^{+0.28}_{-0.26}$ & $7.3^{+3.3}_{-2.8}$ & $0.27^{+0.23}_{-0.19}$ & $0.57^{+0.07}_{-0.09}$ & $1.31^{+1.49}_{-0.85}$ & $1.45^{+1.83}_{-1.04}$ & 0\\[2mm]
    \bottomrule
    \end{tabular}
    \label{tab:sims}
\end{table*}

\begin{figure}
    \centering
    \includegraphics[width=\columnwidth]{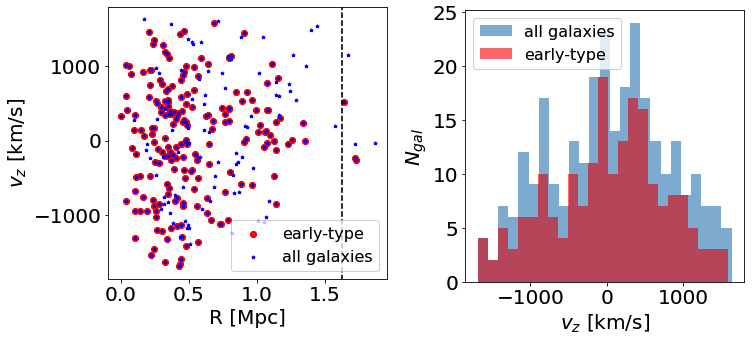}
    \caption{Left: projected-phase-space diagram of Virgo Cluster~A member galaxies. Blue stars indicate the full sample, while red filled circles show the early-type subsample. The vertical dashed line marks the best-fitting value of $r_{200}$ from the \textsc{MG-MAMPOSSt} fit for our reference NFW+pHer+gOM model. Right: LOS velocity distributions of the full sample (blue) and the early-type subsample (red).}
    \label{fig:vlos}
\end{figure}

The reconstruction of the orbit of NGC~4569 requires an estimate of the mass profile of Cluster~A. We derive it using the \textsc{MG-MAMPOSSt} code \citep{Pizzuti:2020tdl,Pizzuti:2022ynt}, an extension of the \textsc{MAMPOSSt} algorithm \citep{Mamon2013} that simultaneously constrains the total gravitational potential and orbital-anisotropy profile of galaxy clusters from the kinematics of their member galaxies.

\textsc{MAMPOSSt} (Modelling Anisotropy and Mass Profile of Spherical Observed Systems) assumes spherical symmetry and treats galaxies as collisionless tracers of the gravitational potential. Under these assumptions, the dynamics are governed by the spherical Jeans equation,
\begin{equation}\label{eq:jeans}
\frac{\mathrm{d}(\nu\sigma_r^2)}{\mathrm{d}r}
+2\beta(r)\frac{\nu\sigma_r^2}{r}
=-\nu(r)\frac{\mathrm{d}\phi}{\mathrm{d}r}\,.
\end{equation}
Here, $r$ is the three-dimensional radial distance from the cluster centre, $\nu(r)$ is the tracer number-density profile, $\sigma_r$ is the radial velocity dispersion, and $\phi$ is the total gravitational potential. The velocity anisotropy is defined as $\beta\equiv1-(\sigma_\theta^2+\sigma_\varphi^2)/(2\sigma_r^2)$, where $\sigma_\theta^2$ and $\sigma_\varphi^2$ are the tangential and azimuthal velocity-dispersion components, respectively. In spherical symmetry, $\sigma_\theta^2=\sigma_\varphi^2$, and the anisotropy profile simplifies to $\beta=1-\sigma_\theta^2/\sigma_r^2$.

\textsc{MAMPOSSt} and \textsc{MG-MAMPOSSt} take as input parametric forms for the gravitational potential, tracer number-density profile, and velocity-anisotropy profile. A maximum-likelihood analysis then determines the parameter set that best describes the distribution of cluster members in projected phase space, $(R,v^\parallel)$. Here, $R$ is the cluster-centric distance projected on the sky and $v^\parallel$ is the galaxy LOS velocity in the cluster rest frame.

We denote by $\boldsymbol{\theta}_\nu$, $\boldsymbol{\theta}_b$, and $\boldsymbol{\theta}_m$ the parameter vectors describing the tracer number-density profile, velocity anisotropy, and gravitational potential, respectively. For each parameter set $\boldsymbol{\Theta}=\{\boldsymbol{\theta}_\nu,\boldsymbol{\theta}_b,\boldsymbol{\theta}_m\}$, the code computes the sum of the log-probabilities $\ln q(R_i,v_i^\parallel\mid\boldsymbol{\Theta})$ of finding a galaxy at the projected-phase-space position $(R_i,v_i^\parallel)$. A detailed description of the algorithm is given by \citet{Mamon2013}; here we briefly summarise the main steps used to obtain $\ln q$:
\begin{itemize}
    \item Given $\boldsymbol{\Theta}$, the Jeans equation (Eq.~\ref{eq:jeans}) is solved to obtain $\sigma_r^2(r)$.
    \item Assuming a functional form for the three-dimensional velocity distribution $f(\boldsymbol{v})$, \textsc{MAMPOSSt} integrates over the two velocity components perpendicular to the LOS to obtain the LOS velocity distribution $h(v^\parallel,R,r)$.
    \item The probability $q$ is then obtained by integrating $h(v^\parallel,R,r)$ along $r$, weighted by the projected tracer surface density $\Sigma(R)$.
\end{itemize}
Our version of \textsc{MG-MAMPOSSt} assumes a Gaussian three-dimensional velocity distribution. In this case, $h(v^\parallel,R,r)$ is also Gaussian, with velocity dispersion
\begin{equation}
\sigma_\parallel^2(R,r)
=\left[1-\beta(r)\left(\frac{R}{r}\right)^2\right]\sigma_r^2(r)\,.
\end{equation}

\textsc{MG-MAMPOSSt} extends the models available in \textsc{MAMPOSSt} to additional parametrisations of the gravitational potential, including modified-gravity and dark-energy scenarios that are viable at cosmological scales (see \citealp{Pizzuti:2020tdl}), and to general velocity-anisotropy profiles that encompass a wide range of member-galaxy orbital distributions.
A recent version of the code also includes a module that incorporates velocity-dispersion measurements of the brightest cluster galaxy (BCG) into the fit, providing tighter constraints on the slope of cluster dark-matter profiles (see \citealt{Biviano:2023oyf}).

The Virgo cluster as a whole is not dynamically relaxed (see e.g. \citealt{Schindler99,Conselice2001,Boselli14c}). Since the Jeans analysis assumes that galaxies are collisionless tracers of the potential in dynamical equilibrium, departures from equilibrium may introduce systematic effects in the mass determination. This bias is particularly important when reconstructing the mass profile in modified gravity \citep{Pizzuti2020}. However, the tests performed by \citet{Mamon2013} showed that \textsc{MAMPOSSt} can recover the virial mass $M_{200}$ reasonably well in standard gravity, even for unrelaxed systems and velocity distributions that deviate from the Gaussian assumption. In addition, we restrict our analysis to Virgo Cluster~A, the subsystem that dominates the diffuse X-ray emission of Virgo \citep{Schindler99}. Previous studies indicate that Cluster~A is close to dynamical equilibrium when traced by its early-type galaxy population \citep{Conselice2001,Kashibadze2020}. The right panel of Fig.~\ref{fig:vlos} compares the LOS velocity distributions of the full member catalogue and the early-type subsample.

In this work, we apply \textsc{MG-MAMPOSSt} to infer the dynamical mass profile of Cluster~A using two datasets: the full member catalogue and a subsample restricted to early-type galaxies. The full sample contains 285 cluster members, selected as described in Sect.~\ref{sec:Virgo}, within a projected radius of $1.8\,\mathrm{Mpc}$ from the cluster centre, identified with the position of M87. Of these, 185 are classified as early-type galaxies. Figure~\ref{fig:vlos} shows the projected phase space (left) and LOS velocity distributions (right) of the full sample over $R\in[0,1.9]\,\mathrm{Mpc}$, in blue, and of the early-type subsample, in red.

In principle, \textsc{MG-MAMPOSSt} can jointly fit the three model components $\nu(r,\boldsymbol{\theta}_\nu)$, $\beta(r,\boldsymbol{\theta}_b)$, and $\phi(r,\boldsymbol{\theta}_m)$ in a parameter space of up to 10 dimensions. However, the galaxy number-density profile $\nu(r)$ can be constrained independently from the projected phase-space data using the unbinned method of \citet{Sarazin80}. We consider two models for the projected tracer profile: a projected NFW profile (pNFW) and a projected Hernquist profile (pHer; see below). Each depends on a scale radius $r_\nu$ and a normalisation. Since the normalisation factors out of the expression for $\sigma_r^2(r)$, only $r_\nu$ enters the \textsc{MG-MAMPOSSt} analysis.

For the cluster gravitational potential, we consider three mass-profile models: the NFW profile, adopted as our reference model; the Hernquist profile \citep{Hernquist01},
\begin{equation}
M_\mathrm{Her}(r)
=M_{200}\frac{(r_{200}+r_\mathrm{s})^2}{r_{200}^2}
\frac{r^2}{(r+r_\mathrm{s})^2}\,,
\end{equation}
and the Burkert profile \citep{Burkert01},
\begin{equation}
\begin{split}
M_\mathrm{Bur}(r)=
\frac{
M_{200}\left[
\ln\left((r/r_\mathrm{s})^2+1\right)
+2\ln(r/r_\mathrm{s}+1)
-2\tan^{-1}(r/r_\mathrm{s})
\right]
}{
\ln\left[(r_{200}/r_\mathrm{s})^2+1\right]
+2\ln(r_{200}/r_\mathrm{s}+1)
-2\tan^{-1}(r_{200}/r_\mathrm{s})
}\,.
\end{split}
\end{equation}
All three models are described by two free parameters: the mass $M_{200}$ (or, equivalently, the radius $r_{200}$) and the scale radius $r_\mathrm{s}$, both fitted by \textsc{MG-MAMPOSSt}. Note that $r_\mathrm{s}$ does not have the same meaning for the different mass-profile parametrisations. For the NFW model, \(r_\mathrm{s}\equiv R_\mathrm{s}=r_{-2}\), where \(R_\mathrm{s}\) is the notation used in Sect.~3.4 and \(r_{-2}\) is the radius at which the logarithmic slope of the density profile equals \(-2\). For the Hernquist and Burkert
models, \(r_\mathrm{s}=2r_{-2}\) and \(r_\mathrm{s}=(2/3)r_{-2}\), respectively.

Finally, we adopt two parametrisations for $\beta(r)$ (see, e.g., \citealt{Mamon19}): a generalised Tiret profile \citep{Tiret2007}, hereafter gT,
\begin{equation}
\beta
=\beta_0+(\beta_\infty-\beta_0)
\frac{r}{r+r_\beta}\,,
\label{e:tiret}
\end{equation}
and a generalised Osipkov--Merritt model \citep{Osipkov79,Merritt85}, hereafter gOM,
\begin{equation}
\beta
=\beta_0+(\beta_\infty-\beta_0)
\frac{r^2}{r^2+r_\beta^2}\,.
\label{e:gOM}
\end{equation}
Here, $\beta_0$ and $\beta_\infty$ represent the inner and outer values of the anisotropy, respectively, and $r_\beta$ is a characteristic scale radius. In both cases, we set $r_\beta=r_{-2}$ of the mass profile, as suggested by numerical simulations \citep{mamon10}. In \textsc{MG-MAMPOSSt}, we work with the rescaled parameters $\mathcal{A}_0=(1-\beta_0)^{-1/2}$ and $\mathcal{A}_\infty=(1-\beta_\infty)^{-1/2}$, which are equal to unity for isotropic orbits.

\subsection{Results}

\begin{figure}
    \centering
    \includegraphics[width=0.8\columnwidth]{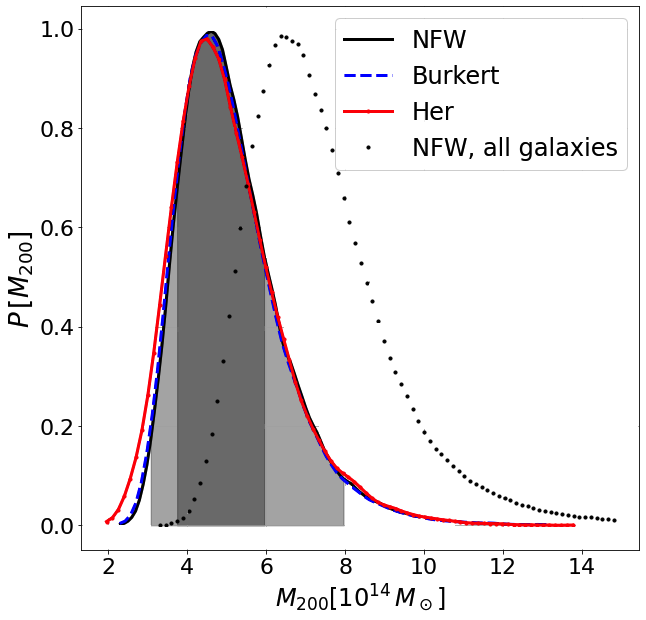}
    \caption{Marginalised distributions of $M_{200}$ obtained from the \textsc{MG-MAMPOSSt} analyses for different mass models. In all cases, a pHer tracer number-density profile and a gOM velocity-anisotropy profile are assumed. The black solid line shows the reference NFW model fitted to the ET sample. Darker and lighter shaded regions indicate the 68\,\% and 95\,\% confidence levels, respectively. The red solid curve shows the Hernquist model for the ET sample, the blue dashed curve the Burkert model for the ET sample, and the black dotted curve the NFW model fitted to the full galaxy sample.}
    \label{fig:m200}
\end{figure}

To characterise the total mass profile of Cluster~A, we examine the combinations of $M(r)$, $\nu(r)$, and $\beta(r)$ models described above. We also test the dependence of the inferred mass profile on tracer selection by analysing both the full galaxy sample and the early-type subsample. We sample the parameter space $\boldsymbol{\Theta}=\{r_{200},r_\mathrm{s},r_\nu,\mathcal{A}_\infty,\mathcal{A}_0\}$ using an MCMC chain of $110\,000$ steps. We discard the first $10\,000$ steps as burn-in and use the remaining $100\,000$ samples for posterior inference, adopting flat, uninformative priors for all parameters. The number-density scale radius $r_\nu$ can be constrained independently of \textsc{MG-MAMPOSSt}, so we first fit the projected galaxy distribution $R_i$ assuming the pHer and pNFW models. We then allow $r_\nu$ to vary in the \textsc{MG-MAMPOSSt} analysis, adopting the $2\sigma$ interval from this preliminary fit as its prior range.

Our results are summarised in Table~\ref{tab:sims}, where we report the constraints on the free parameters for each combination of mass, number-density, and anisotropy models. The last column reports the Bayesian Information Criterion difference (BIC; see e.g. \citealt{Mamon19}) relative to the best-fitting model for each sample. A difference $\Delta\mathrm{BIC}>6$ is generally taken as strong evidence against the model with the larger BIC.

The pNFW tracer number-density model is strongly disfavoured relative to the pHer profile. For the early-type sample, the Hernquist mass model combined with the gOM anisotropy profile gives the lowest BIC. However, the differences relative to the NFW and Burkert mass profiles are not statistically significant ($\Delta\mathrm{BIC}<1$). We therefore retain the NFW profile as our reference mass model.

Figure~\ref{fig:m200} shows the marginalised distributions of $M_{200}=[100H^2(z)/G]r_{200}^3$ for the three mass profiles fitted to the early-type sample, together with the corresponding distribution for the reference NFW model fitted to the full galaxy sample (dotted curve). In all cases, a pHer tracer number-density profile and a gOM anisotropy profile are assumed.

The constraints on \(M_{200}\) obtained with the three mass
models agree closely, indicating limited sensitivity of the
inferred cluster mass to the tested mass-profile parametrisations. For the NFW model, we obtain
\begin{equation}
M_{200}
=5.4^{+2.9}_{-2.3}\times10^{14}\,\mathrm{M}_\odot\,,
\end{equation}
at $2\sigma$, in agreement with \citet{Kashibadze2020}. When the full sample (ET+LT) is used, the NFW mass inferred by \textsc{MG-MAMPOSSt} is larger, $M_{200}=7.5^{+3.8}_{-3.0}\times10^{14}\,\mathrm{M}_\odot$. This suggests that the late-type population may be farther from dynamical equilibrium than the lenticular and elliptical populations, as discussed above. The \textsc{MG-MAMPOSSt} best-fitting values for the full-sample analysis are listed in the last three rows of Table~\ref{tab:sims}.

\section{Additional orbital diagnostics}
\label{app:orbital_diagnostics}

\begin{figure}
\centering
\includegraphics[width=\columnwidth]
{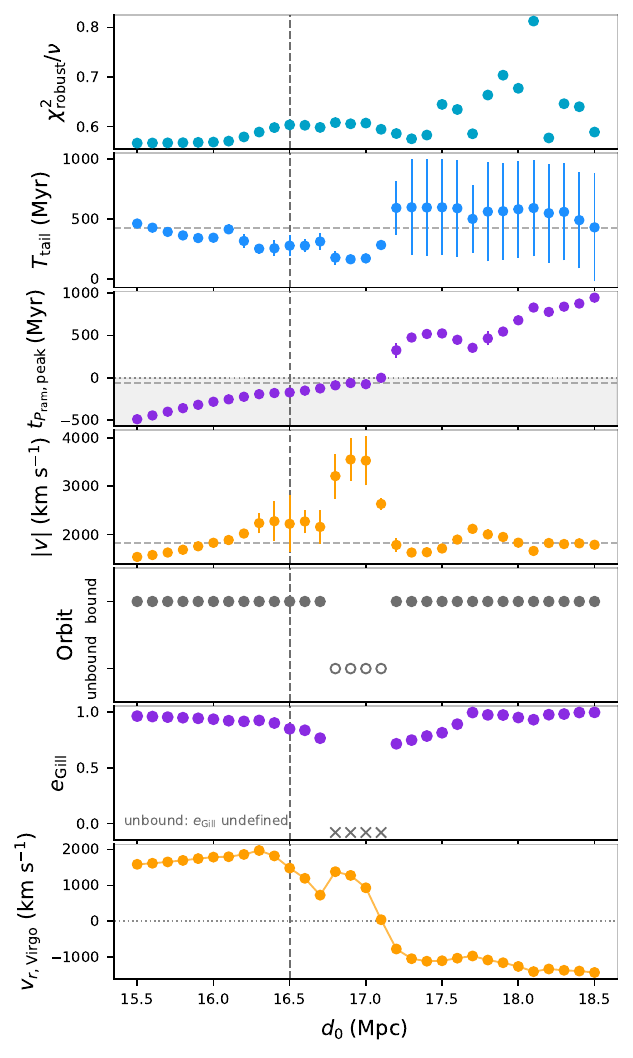}
\caption{
Extended orbital diagnostics across the LOS-distance scan.
The first five panels reproduce the quantities shown in
Fig.~\ref{fig:distance_scan_continuity}. The sixth panel shows the eccentricity \(e=1-r_{\rm peri}/r_{\rm apo}\) for bound solutions; grey crosses mark distances with unbound solutions,
for which the eccentricity is undefined. The bottom panel shows the present-day cluster-centric radial velocity, with positive and negative values indicating outward and inward motion, respectively.
The horizontal dotted line marks zero radial velocity,
and the vertical dashed line marks the fiducial distance
\(d_0=16.5\,\mathrm{Mpc}\).
}
\label{fig:orbital_diagnostics}
\end{figure}

We complement the distance scan in
Sect.~\ref{sec:distance_scan} with two diagnostics: the orbital eccentricity and the present-day cluster-centric
radial velocity. Figure~\ref{fig:orbital_diagnostics}
shows their dependence on the assumed LOS distance. 

Following \citet{Gill2004}, we define the eccentricity as \(e\equiv1-r_{\rm peri}/r_{\rm apo}\), where \(r_{\rm peri}\) and \(r_{\rm apo}\) are the turning radii of the reconstructed galaxy orbit in the adopted cluster potential. This definition applies to bound solutions;
unbound orbits have no apocentre.
Solutions at \(d_0\simeq16.5\)--\(16.7\,\mathrm{Mpc}\) and
\(17.2\)--\(17.6\,\mathrm{Mpc}\), including the fiducial case, have \(e\simeq0.7\)--\(0.9\). For the first-passage interpretation considered here, the relevant comparison is the recently infalling satellite population identified by \citet{Gill2004}, for which they report a mean eccentricity of approximately \(0.8\).
Their simulations also show that eccentricity tends to decrease with successive orbits. The reconstructed eccentricities in these distance intervals are therefore comparable to those of recently infalling satellites. Other bound solutions approach unity, implying more extreme apocentre-to-pericentre ratios. This comparison complements the boundedness diagnostic but does not select a unique distance.

The present-day cluster-centric radial velocity is 
\begin{equation}
v_{r,\mathrm{cl}}(0)
=
\left.\frac{\mathrm{d}D_\mathrm{cluster}}{\mathrm{d}t}
\right|_{t=0}
=
\frac{\mathbf{R}_0\cdot\mathbf{V}_0}
{|\mathbf{R}_0|},
\label{eq:cluster_radial_velocity}
\end{equation}
where \(\mathbf{R}_0\) and \(\mathbf{V}_0\) are the present-day galaxy position and velocity relative to the Virgo Cluster~A centre. The low-distance solutions have
\(v_{r,\mathrm{cl}}(0)>0\), corresponding to outward motion after pericentre, whereas the high-distance solutions have \(v_{r,\mathrm{cl}}(0)<0\), corresponding to inward motion before pericentre. The zero crossing near \(d_0\simeq17.1\,\mathrm{Mpc}\) coincides with the reconstructed ram-pressure maximum crossing the present epoch. The broader transition interval
\(d_0\simeq16.8\)--\(17.1\,\mathrm{Mpc}\) encompasses the accompanying changes in the fitted orbital parameters and boundedness; only near its upper edge is the galaxy reconstructed close to pericentre at the present epoch. Each distance corresponds to a separately fitted orbit.

\section{Wind--disc geometry}
\label{app:wind_disc_angle}
\begin{figure}
    \centering
    \includegraphics[width=\columnwidth]
    {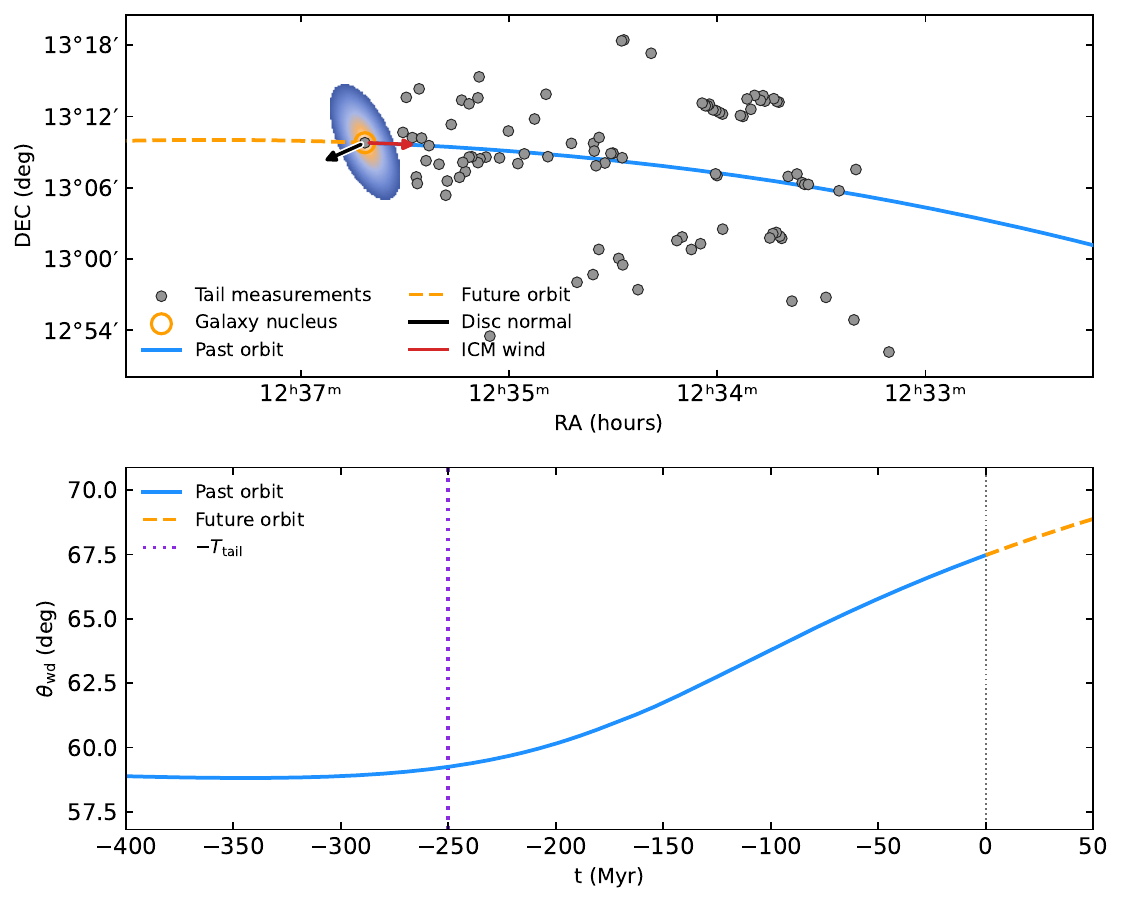}
    \caption{Wind--disc geometry for the fiducial
    \(d_0=16.5\,\mathrm{Mpc}\) reconstruction. \emph{Top panel}: projected orientation of the NGC~4569 disc and reconstructed orbit at the present epoch. The black and red arrows indicate the disc normal  and the ICM flow velocity in the galaxy rest frame, respectively. Their apparent projected separation does not correspond to \(\theta_{\rm wd}\), which is evaluated from their full three-dimensional orientations. \emph{Bottom panel}: evolution of the wind--disc angle along the reconstructed past and forward-integrated future orbit.}
    \label{fig:wind_disc_angle}
\end{figure}
\begin{figure}
    \centering
    \includegraphics[width=\columnwidth]{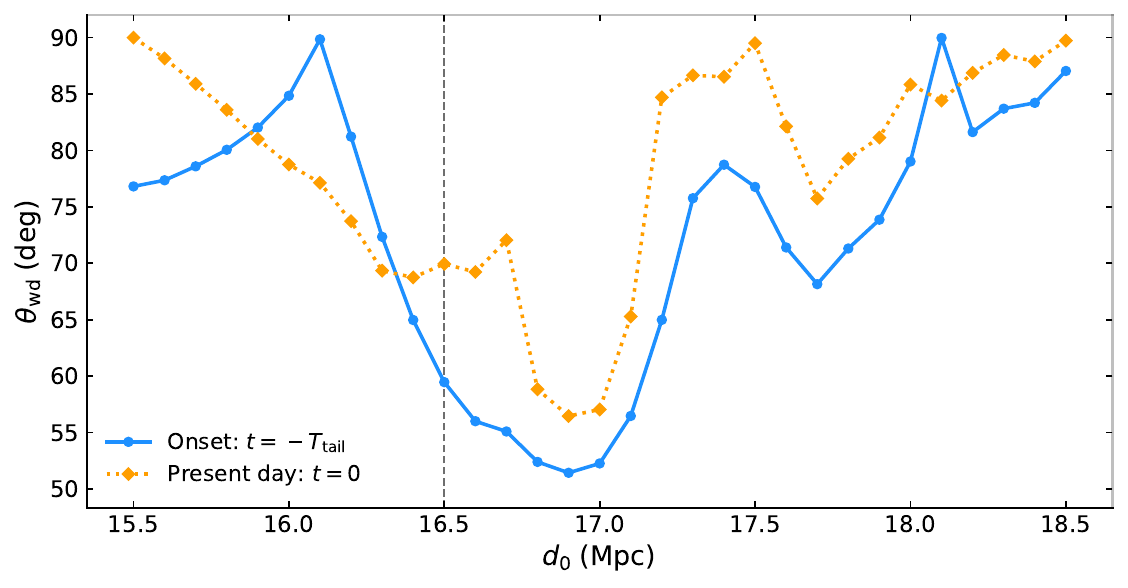}
    \caption{Wind--disc angle as a function of the assumed
LOS distance of NGC~4569. The curves show \(\theta_{\rm wd}\) at the
lookback time assigned to the oldest fitted H\(\alpha\) material and at
the present epoch, derived from the best-fitting orbit at each distance.
The vertical dashed line marks the fiducial distance
\(d_0=16.5\,\mathrm{Mpc}\). Angles of \(0^\circ\) and \(90^\circ\)
correspond to face-on and edge-on interactions, respectively.}\label{fig:wind-disk_distance}
\end{figure}
The reconstructed orbit also constrains the orientation
of the ICM wind relative to the disc of NGC~4569. We construct the
disc-normal vector, \(\mathbf{n}_{\rm disc}\), in the ICRS frame by
adopting a disc inclination to the LOS \(i=65^\circ\), corresponding to the average
of the available photometric and kinematic estimates
\citep{Haynes84,Binggeli85,Chemin2006}, and a position angle
\(\mathrm{PA}=24^\circ\) \citep{Vollmer04}. The disc orientation is
assumed to remain fixed over the reconstructed interval. Neglecting
bulk motions of the ICM, the wind velocity in the galaxy rest frame is
\(\mathbf{W}(t)=-\mathbf{v}_{\rm gal}(t)\). We define the wind--disc
angle as

\begin{equation}
\theta_{\rm wd}(t)=
\arccos\left[
\frac{\left|\mathbf{W}(t)\cdot\mathbf{n}_{\rm disc}\right|}
{|\mathbf{W}(t)|\,|\mathbf{n}_{\rm disc}|}
\right],
\label{eq:wind_disc_angle}
\end{equation}
where the absolute value accounts for the twofold
orientation ambiguity of the disc normal. With this convention,
\(\theta_{\rm wd}=0^\circ\) corresponds to a face-on interaction and
\(\theta_{\rm wd}=90^\circ\) to an edge-on interaction.

Figure~\ref{fig:wind_disc_angle} shows the geometry
of the fiducial reconstruction at \(d_0=16.5\,\mathrm{Mpc}\). During
the interval sampled by the fitted H\(\alpha\) tail,
\(\theta_{\rm wd}\) increases from
\(\theta_{\rm wd}(-T_{\rm tail})=59.3^\circ\) to
\(\theta_{\rm wd}(0)=67.5^\circ\). The interaction therefore remains oblique, but closer to edge-on than
face-on, throughout the reconstructed tail history. This is broadly consistent with the
inclined stripping configuration adopted by \citet{Vollmer04}.

The dependence on the assumed LOS distance is shown in
Fig.~\ref{fig:wind-disk_distance}, where \(\theta_{\rm wd}\) is
evaluated both at the lookback time assigned to the oldest fitted
H\(\alpha\) material and at the present epoch. Although the detailed
angle varies with \(d_0\), particularly across the transition between
the two orbital branches, it remains above approximately \(50^\circ\)
throughout the explored range. The interaction is therefore
consistently reconstructed as oblique and closer to edge-on. The
distance dependence of \(\theta_{\rm wd}\) is not used as an
independent distance constraint, but provides the wind geometry
associated with each reconstructed orbit for future tailored
hydrodynamical simulations.

\end{document}